\documentclass[iop]{emulateapj}
\usepackage{apjfonts}
\pdfoutput=1

\usepackage[bookmarks=true, colorlinks=true, citecolor=blue,backref=true]{hyperref}

\newcommand{\ha}{$\rm{H}\alpha$}

\newcommand{\mic}{\mu{\rm m}}

\newcommand{\galex}{{\it GALEX}}
\newcommand{\wise}{{\it WISE}}
\newcommand{\her}{{\it Herschel}}

\slugcomment{}

\shorttitle{IRX-$\beta$ RELATION}
\shortauthors{SALIM \& BOQUIEN}

\begin{document}

\title{Diversity of Galaxy Dust Attenuation Curves Drives the Scatter in
  the IRX-$\beta$ Relation}

\author{Samir Salim\altaffilmark{1}, M\'ed\'eric
  Boquien\altaffilmark{2}}
\altaffiltext{1}{Department of Astronomy, Indiana University,
   Bloomington, IN 47404, USA} 
\altaffiltext{2}{Centro de Astronom\'ia (CITEVA), Universidad de Antofagasta, Avenida Angamos 601, Antofagasta, Chile}
\email{salims@indiana.edu}

\begin{abstract}
  We study the drivers of the scatter in the IRX-$\beta$ relation
  using 23,000 low-redshift galaxies from the \galex-SDSS-\wise\
  Legacy Catalog 2 (GSWLC-2). For each galaxy we derive, using CIGALE
  and the SED+LIR fitting technique, the slope of the dust attenuation
  curve and the strength of the UV bump, plus many other galaxy
  parameters. We find that the IRX-$\beta$ scatter is driven entirely
  by a wide range of attenuation curves--- primarily their
  slopes. Once the slope and the UV bump are fixed, the scatter in the
  IRX-$\beta$ vanishes. The question of the IRX-$\beta$ scatter is the
  direct manifestation of a more fundamental question of the diversity
  of dust attenuation curves. The predominant role of the attenuation
  curve is the consequence of a narrow range of intrinsic UV slopes of
  star-forming galaxies. Galaxies with different specific SFRs or
  population ages do not show strong trends in the IRX-$\beta$ diagram
  because their attenuation curves are, on average,
  similar. Similarly, there is no shift in the IRX-$\beta$ locus
  between starbursts and normal star-forming galaxies, both types
  having, on average, steep attenuation curves. Optical opacity is
  identified as the strongest determinant of the attenuation curve
  slope, and consequently the IRX-$\beta$ diversity. Despite the
  scatter, the use of an average IRX-$\beta$ relation is justified to
  correct SFRs, adding a random error of $\lesssim0.15$ dex. The form
  of the local correspondence between IRX-$\beta$ and attenuation
  curves is maintained at high redshift as long as the evolution of
  the intrinsic UV slopes stays within a few tenths.

\end{abstract}


\section{Introduction} \label{sec:intro}

Determination of a galaxy's star formation rate (SFR) is a critical
task for the study of galaxy evolution. Many SFR indicators have been
proposed and utilized over the years \citep{kennicutt98,kennicutt12},
but two classes of methods stand out as the most practical over a wide
range of redshifts: 1) SFRs based on the nebular line emission
\citep{kennicutt83} and 2) SFRs based on the stellar continuum emission
\citep{larson78}. However, both classes of methods are critically
limited by our ability to correct for the effects of dust, which
require both the knowledge of the dust attenuation law and the
knowledge of the intrinsic unattenuated spectral energy distribution
(SED). These limitations have been already recognized in  
\citet{pettini98}, and have not yet been fully overcome.

Line emission method typically utilizes H$\alpha$ luminosity (less
frequently [OII]3727) of very young stars ($<10$ Myr) still embedded
in molecular clouds and HII regions.  The H$\alpha$ luminosity must be
corrected for dust emission, which is usually accomplished by
observing H$\beta$, which gives the correction to H$\alpha$ luminosity
via the Balmer decrement plus an appropriate attenuation or extinction
curve. Many studies use Milky Way extinction curve (e.g.,
\citealt{cardelli89}) to correct for the dust affecting emission
lines. The correction will depend on the relative attenuation curve
between H$\beta$ and H$\alpha$ wavelengths, i.e., it is not sensitive
to the shape of the assumed curve at shorter wavelengths. Getting both
H$\alpha$ and H$\beta$ requires spectroscopic observations, which
limits the samples with available data. Pa$\alpha$, an emission line
in the near-IR, is much less affected by the dust and represents a
better alternative to \ha\ (e.g., \citealt{calzetti07}), but the
observed samples are relatively small. Furthermore, emission line
fluxes need to be corrected for aperture (fiber) or slit losses.

There are several closely related methods that use continuum emission
to derive SFR, either arising from direct stellar emission, or from
the emission of dust heated by relatively young ($<100$ Myr)
stars. Traditionally, the continuum methods are implemented as: (a)
the ultraviolet (UV) method, (b) the infrared (IR) method, (c) the
UV+IR method, (d) the SED fitting (stellar emission) and (e) the
energy-balance SED fitting (stellar plus dust emission). The UV method
obtains the SFR from the far-UV (FUV) luminosity and the UV color,
from which the dust correction in the FUV is derived by assuming some
correlation between FUV attenuation ($A_{\rm FUV}$) and the UV
color. For star-forming galaxies, the FUV attenuation is directly
related to the ratio of IR luminosity to FUV luminosity, i.e., the IR
excess (IRX, \citealt{meurer99,gordon00}). Thus the accuracy of the UV
method is predicated on there being a relatively tight and universal
relation between IRX and UV color, or, alternatively, IRX and the UV
spectral slope ($\beta$;
$f_{\lambda}(\lambda)\propto \lambda^{\beta}$), the so called
IRX-$\beta$ relation \citep{meurer95,meurer99}. Correcting the
rest-frame UV luminosities with the UV method is employed at high
redshift where IR information is very limited or non existent (e.g.,
\citealt{reddy09,bouwens09}).

Next, in the IR method, the estimate of the SFR is based on the light
absorbed by the dust and reemitted in the thermal IR. It assumes that
all emission from young stars is processed by dust and that the
contribution of older populations is small \citep{kennicutt98}. Its
refinement, the UV+IR method (e.g., \citealt{elbaz07,daddi07,reddy12})
combines the observed (attenuated) SFR in the UV with the obscured SFR
in the IR, but still requires dust heating by older stars to be small
in order not to bias the answer, or else, it needs to be accounted for
\citep{buat11,boquien16}. Standard SED fitting \citep{conroy13} of
just the stellar emission (e.g., \citealt{papovich01,shapley05,s07})
is essentially the unabridged version of the plain UV method, but with
multiple bands constraining the SFR. The SED fitting essentially
derives dust-corrected SFRs also via a relation between UV color and
FUV attenuation, but the relation is present implicitly in the models
and is typically implemented via dust attenuation curve(s) and not an
explicit IRX-$\beta$ relation. Finally, the energy-balance SED fitting
\citep{burgarella05,dacunha08} directly includes IR in constraining
the SFR, which also helps to constrain the parameters of the
attenuation curve (assuming it was not fixed in the fitting), while
correctly attributing the IR emission to dust heating from both young
and old stars.
\citep{burgarella05,noll09,boquien12,boquien16,leja17,s18}.

In the absence of IR data, many high-redshift studies rely on the UV
method, and thus need to assume some IRX-$\beta$ relation, or,
equivalently, an $A_{\rm FUV}$-$\beta$ relation. The relationship
between the FUV attenuation (or IR excess) and the UV spectral slope
was originally studied based on $\sim 60$ relatively compact local
starbursts (and blue compact dwarfs) by \citet{meurer99}, drawn from
the UV spectral atlas of \citet{kinney93}, produced with {\it
  International Ultraviolet Explorer (IUE)}, and with IR luminosities
derived from {\it Infrared Astronomical Satellite (IRAS)}
observations.  \citet{meurer99} found that these starbursts obey a
relatively tight IRX-$\beta$ relation. The existence of a relation
lend credence to using applying it to higher-redshift samples
($z\gtrsim 1$), especially because the local starbursts may share many
of the characteristics of young galaxies at high redshift.

The possibility that the IRX-$\beta$ relation could be universal and
relatively tight for non-starbursting galaxies was challenged by
\citet{kong04}, who found that including normal galaxies (observed by
{\it IUE} and several other early UV satellites) introduced an
additional scatter and a possible shift with respect to the starburst
relation of \citet{meurer99}. Moreover, \citet{kong04} have suggested
that the offset and the additional scatter may be due to a wide range
of older stellar populations found in normal star-forming galaxies,
with offsets correlating with the stellar population age.

Possible non-universality of the IRX-$\beta$ relation was indicated in
numerous subsequent studies (e.g.,
\citealt{buat05,seibert05,gildepaz07,dale09,takeuchi10,grasha13}),
which utilized much larger sets of observations of more normal
galaxies acquired by {\it Galaxy Evolution Explorer (GALEX)} in its
two UV bands (FUV and NUV), and compared the resulting IRX-$\beta$
distributions of points to the starburst relations of \citet{meurer99}
and \citet{kong04}, finding offsets.  However, remeasuring of
\citet{meurer99} galaxies using {\it GALEX} revealed that {\it IUE}
missed a large fraction of FUV emission
\citep{overzier11,takeuchi12,casey14} biasing the original starburst
relation with respect to the one obtained by \galex for the same
starburst galaxies. Nevertheless, the question of the drivers of the
large scatter in the IRX-$\beta$ relation (the so called second
parameter in that relation) remained.

Following \citet{kong04}, much attention was paid on investigating the
role of the age, or age-related metrics, on the dispersion in the
IRX-$\beta$, both empirically \citep{burgarella05,seibert05,panuzzo07,
  johnson07,boquien09,grasha13} and using radiative transfer modeling
\citep{popping17,safarzadeh17,narayanan18}. Factors other than the age
were explored as well, such as the variations in the intrinsic UV
slope \citep{boquien12} and the role of the dust type and/or the
geometry \citealt{witt00,bell02,inoue06,thilker07,panuzzo07,popping17,
  safarzadeh17,narayanan18}. Both the dust type and the geometry
affect the resulting attenuation curve--the attenuation as a function
of wavelength normalized by attenuation at some wavelength, usually
$V$ band.

The role of the attenuation curve in driving the scatter in the
IRX-$\beta$ plane has been explicitly studied in
\citet{burgarella05,boquien09, boquien12,buat12}, or is sometimes
implicitly assumed to be a factor, especially in high-redshift studies
(e.g., \citealt{meurer95,siana09,salmon16,bouwens16,
  reddy18,cullen17,mclure18}). However, its precise role is difficult
to establish because measuring the attenuation curve for individual
galaxies represents a major challenge (e.g.,
\citealt{kriek13,salmon16,leja17,s18,buat18}.

IR observations from {\it Spitzer} allowed the study of the
IRX-$\beta$ relation to be carried out out to $z \gtrsim 1$, for IR
luminous galaxies \citep{noll09b,reddy10,murphy11,buat12} and strongly
lensed sources \citep{siana09,pope17}, again suggesting a large spread
in the IRX-$\beta$ plane. These redshifts were subsequently explored
at longer wavelengths with {\it Herschel} \citep{nordon13}. At even
higher redshifts ($z \gtrsim 3$), some studies using {\it Herschel}
and ALMA suggest a stronger evolution in the IRX-$\beta$ relation
\citep{capak15,bouwens16,barisic17}, but the question of how IR
luminosity is measured start to arise as well
\citep{lee12,faisst17,cullen17,ferrara17}. Recently, efforts were made
to derive an average IRX-$\beta$ relation for general population of
high-redshift ($z\gtrsim 2$) galaxies using the stacking of {\it
  Herschel}, SCUBA-2 or ALMA observations at longer wavelengths (e.g.,
\citealt{heinis13,pannella15,bouwens16,forrest16,alvarez16,bourne17,fudamoto17,
  reddy18,mclure18,koprowski18}), but the question of the drivers of
the scatter, which appears to be present at all redshifts, remains a
fundamental one.

Our approach in addressing the issue of the IRX-$\beta$ scatter is
that using very large samples of local galaxies with robustly
determined galaxy parameters can help inform the underlying mechanisms
that may be valid regardless of the redshift. Recently, we have
constructed \galex-SDSS-\wise\ Legacy Catalog (GSWLC, \citealt{s16}),
a catalog of physical parameters of galaxies obtained using Bayesian
SED fitting. GSWLC is drawn from SDSS spectroscopic survey, with
UV observations from \galex, and mid-IR observations from \wise,
containing altogether 700,000 optically selected galaxies. In a recent
update of the catalog (GSWLC-2, \citealt{s18}), the IR luminosity was
used to constrain the parameters of the dust attenuation curve for
individual galaxies. This important additional information allows us
to systematically explore, in the present work, a range of different
galaxy properties as potential drivers of the IRX-$\beta$ scatter of
entire galaxies.

The paper is organized as follows. Section \ref{sec:sample} provides
the summary of the sample and the data. In Section \ref{sec:method} we
describe the derivation of galaxy parameters, and the methodology for
selecting the fiducial IRX-$\beta$ relation. In Section
\ref{sec:results} we present the results of the systematic
investigation of the drivers of the scatter, using non-parametric
approach. In Section \ref{sec:eta} we introduce new parameterization
useful for studying the trends in the IRX-$\beta$ plane. Recipes that
include extensions of the IRX-$\beta$ relation are presented in
Section \ref{sec:recipes}. The results are discussed in Section
\ref{sec:disc} and summarized in Section \ref{sec:conc}. Throughout
this work we assume WMAP7 flat cosmology ($H_0=70.4$ km s$^{-1}$
Mpc$^{-1}$, $\Omega_m=0.272$).

\section{Sample and data} \label{sec:sample}

In this work we use the deep UV photometry catalog from
GALEX-SDSS-WISE Legacy Catalog 2 (GSWLC-D2). GSWLC is a publicly
available catalog of SED fitting parameters derived by combining
\galex, SDSS and \wise\
photometry.\footnote{\url{http://pages.iu.edu/\~salims/gswlc}} GSWLC-1
\citep{s16} and GSWLC-2 \citep{s18} share the sample, but the latter
performs energy-balance SED+LIR fitting. The GSWLC sample includes all
galaxies from the SDSS DR10 spectroscopic surveys such that
$0.01<z<0.30$ and $r_{\rm petro}<18.0$, as long as they fall within
the \galex\ coverage. Because \galex\ observations have a very wide
range of exposure times, separate samples are defined for shallow
(all-sky; GSWLC-A), medium-deep (GSWLC-M) and deep (GSWLC-D) UV
observations. Details of sample construction and matching are given in
\citet{s16}. 

While the GSWLC-D2 catalog is the smallest of the three, covering 7\%
of SDSS footprint, it has the highest quality of UV photometry, with
the UV exposure time of at least 4000 s, and hence provides the most
reliable UV slope ($\beta$), which is why we choose it as the sample
for this study. GSWLC-D2 contains 48,401 galaxies, which reduces to
47,672 after the exclusion of type 1 AGN (QSO-like spectra) and
objects with poor ($\chi^2_r>30$) SED fits. Next we require that both
the FUV and NUV are detected at $>3\sigma$, which leaves 25,791
objects. This cut mostly removes early-type, passive galaxies, which
are not the focus of the study, considering that their FUV attenuation
does not follow the tight relation with IRX as it does for the
star-forming galaxies \citep{cortese08,viaene16}. Furthermore, we
require mid-IR detection in either 12 $\mic$ or 22 $\mic$, which
results in 23,175 galaxies, which we refer to as ``all'' galaxies. In
addition, we use the \citet{bpt} (BPT) emission-line diagram to select
star-forming (SF) galaxies, as described in \citet{s18}. More
accurately, we exclude 5501 galaxies on the AGN branch and 5585
galaxies with weak lines (S/N of H$\alpha$ flux $<10$) to get the
``star-forming'' (SF) sample of 12,089 galaxies. The star-forming
sample encompasses the main sequence and extends down to log
sSFR$\approx -11$. Selecting the SF sample based on the SDSS fiber
spectroscopy may miss galaxies that have the bulk of their SF outside
of the $3"$ fiber (typically, early-type galaxies in the green valley,
\citealt{s10}), but such galaxies are relatively rare (see Fig.\ 8 in
\citealt{s16}). Alternatively, we could have selected galaxies with
specific SFR (sSFR) above some threshold, regardless of the BPT class,
and this would not have affected the results at all.

We use the \galex\ pipeline photometry with the application of some
corrections, as described in \citet{s18}. For optical photometry we
use SDSS $ugriz$ \texttt{modelMag} measurements. Finally, we use the
unWISE forced photometry of \citet{lang16} for fluxes at 12 $\mic$ and
22 $\mic$.

\section{Methodology} \label{sec:method}

%
\subsection{Derivation of galaxy parameters} \label{ssec:par}

All galaxy parameters employed in this work are derived using Code
Investigating GALaxy Emission (CIGALE, \citealt{noll09,boquien18}) by
applying the SED+LIR fitting, a variant of the energy-balance SED
fitting\citet{s18} . Specifically, we use CIGALE Version 0.11, but
with custom modifications, some of which were implemented in 0.12 and
2018.1 versions of the code. Energy-balance SED fitting
\citep{dacunha08,noll09} requires that the energy absorbed by the dust
in the UV to the near-infrared should match the luminosity emitted by
the dust, i.e., the total IR luminosity, $L_{\rm IR}$. Energy-balance
SED fitting codes typically fit both the stellar and IR SED, with a
number of parameters required to specify both. In the SED+LIR fitting,
the IR luminosity is determined separately, and is then used as a
constraint in the broadband SED fitting, without the need to fit the
IR SED. The main advantage of the SED+LIR method compared to
UV/optical+IR SED fitting is that the model library for the SED
fitting is not inflated by all of the parameter combinations needed to
model the IR SED, allowing for faster fitting and/or allowing for
other parameters, e.g., the attenuation curve, to be left free.

IR luminosities to be used in the SED+LIR fitting are derived in a
two-step process. We first use luminosity-dependent templates of
\citet{ce01} together with 22 $\mic$ flux (or 12 $\mic$, if not
detected at 22 $\mic$) to get an estimate of the IR luminosity. We
these estimates we apply empirically derived corrections (as a
function of $L_{\rm IR}$ and $L_{\rm IR}/M_*$) constructed to
reproduce IR luminosities of galaxies with the far-IR coverage from
the \her-ATLAS survey \citep{valiante16}. This two-step method results
in IR luminosities that have a remarkable accuracy of $\sim 0.1$ dex,
and show no systematic differences with respect to \her\ luminosities
over the entire $8.8<\log L_{\rm IR}<11.8$ range. The
$L_{\rm IR}/M_*$-dependent correction eliminates deviations of
\citet{ce01} templates reported for starbursting galaxies
\citep{overzier11}. Details are given in \citet{s18}. We have also
checked whether the energy balance assumption is valid for different
viewing geometries. We looked at the difference between the observed
IR luminosity and the dust-absorbed luminosity estimated from the SED
fitting without IR constraints, as a function of galaxy inclination,
and find no systematic difference greater than 0.04 dex.

The use of the IR luminosity allows the dust attenuation curve to be
relatively well constrained when its parameters are left free in the
SED fitting. Following \citet{noll09}, the attenuation curve is
parameterized with a power-law slope deviation from the
\citet{calzetti00} curve ($\delta$), to which the UV bump feature is
added having an amplitude $B$. As discussed in \citet{s18}, the
modification of the slope and the addition of the UV bump is carried
out on $E(B-V)$ normalized \citet{calzetti00} curve. Attenuation curve
slopes are allowed to vary from $\delta=-1.2$, which is steeper than
even the SMC extinction curve, to $\delta=0.4$, somewhat shallower
than the Calzetti curve. The amplitude of the UV bump, centered at
2175 \AA, can vary between $B=-2$ and $B=6$, twice as strong as the MW
bump. Since the amplitude of the bump is not constrained with great
accuracy, the inclusion of negative values is meant to offset the
posterior bias \citep{buat12,salmon16}. Typical error on the curve
slope determination is 0.17, and 1.3 on the bump amplitude (both for
the star-forming sample). 

The attenuation curves described above are not effective curves. The
above attenuation curves, of the same slope and the same bump
strength, are applied separately to young and old populations, but
with old population suffering a fraction of the attenuation that
affects young stars still enshrouded in birth clouds, following the
distinction introduced in \citet{cf00}\footnote{Note that the original
  \citealt{cf00} implementation is based on pure power-law attenuation
  curves with no allowance for the UV bump.}. In the nominal run,
using the dust module {\tt dustatten\_calzleit}, this fraction is
fixed at the default value of 0.44, which we confirm to be close to
the average value when the fraction of attenuation affecting older
population is actually allowed to vary.\footnote{Note that the module
  {\tt dustatt\_modified\_starburst}, introduced in CIGALE v2018.1,
  does not contain the age-dependent implementation of attenuation
  curves affecting the stellar continuum in order to accurately
  reproduce the original implementation of \citet{calzetti94} curve as
  an effective curve, but maintaining the ability to modify the slope
  and add the UV bump.} The split between the young and the old
population (i.e., the birth cloud dispersal time) is fixed at 10 Myr,
the value that fits the great majority of galaxies. Assuming the same
intrinsic shapes for attenuation curves of both populations is a
simplification, driven by the lack of a consensus regarding the
appropriate curve for the young population (birth clouds) and the
difficulty of constraining it from the SED fitting
\citep{lofaro17}. Consequently, the results of the study are not
sensitive to the details regarding the curve of the young population.

It should be pointed out that applying the same intrinsic attenuation
curve (same $A_{\lambda}/A_V$) to young and old populations, but with
different normalizations, as we do here, changes the {\it effective}
attenuation curve. Specifically, in our implementation the power-law
exponent of the slope of the effective curve is on average $\sim 0.2$
steeper than the slope of the intrinsic curve, with some 0.1 scatter
around that average. We obtain the above offset by performing the SED
fitting in which the attenuation curve is treated as an effective
curve, without the age-dependent split, and comparing the resulting
slopes with the ones from the age-dependent model. In this paper,
however, when we refer to the slope of the attenuation curve, we mean
the slope deviations from the Calzetti curve implemented as intrinsic
attenuation curves of the two populations. We have found that applying
the modified Calzetti curve without the old/young split, so that the
input and effective curves are one and the same, results in poorer
quality of the fits, presumably because such age-insensitive
application is less physical.

The SED fitting is based on stellar population models of \citet{bc03},
with a Chabrier IMF \citep{chabrier} and four stellar metallicities
(0.004, 0.008, 0.02 and 0.05). We model star-formation histories as
two components (a 10-Gyr old exponentially declining component, and a
more recent burst of varying strength and constant SFR). We include
the contribution of nebular emission lines and nebular
continuum. Details on model assumptions and SED fitting are given in
\citet{s16,s18}.

\subsection{IRX-$\beta$ parameters and the fiducial relation} \label{ssec:irx}

The IRX is defined as the ratio of the total IR luminosity to the
rest-frame FUV luminosity in \galex\ bandpass.\footnote{Furthermore,
  IRX is defined as just the ratio, following
  \citet{meurer99}. Originally, in \citet{meurer95}, IRX was defined
  as the logarithm of the ratio. Note, however, that both of these
  studies define IRX using the far-IR luminosity, unlike most
  subsequent studies that use total IR luminosity.} We determine IRX
from the SED fitting, with the typical uncertainty of 0.10 dex. We
have verified that the models fully cover the parameter space of
observations and that no systematic biases are present between model
and observed values. The UV slope is defined following the original
methodology of \citet{calzetti94}, as the linear fit to rest-frame
fluxes from 10 windows spanning the range from 1268 to 2580 \AA. The
windows were selected to avoid absorption features that may affect the
derivation of the slope, as well as to exclude the UV bump, or at
least its central portion (1950-2400 \AA). Typical error of the UV
slope, determined via the SED fitting that includes the nebular
continuum contribution, is 0.14. We designate the UV slope determined
from spectral windows as $\beta_{\rm C94}$, to differentiate it from
slopes derived in other ways, for example, from \galex\ FUV and NUV
photometry ($\beta_{\rm GLX}\equiv 2.29({\rm FUV}-{\rm NUV})-2$),
which, in the case of low-redshift studies is often based on observed
(non rest-frame) magnitudes. From our sample we get the relationship
between the two definitions of the slope to be, depending on the
redshift:

\begin{eqnarray}
\beta_{\rm GLX} &= &0.17+1.01 \beta_{\rm C94}\qquad (z<0.3), \label{eqn:beta1}\\
\beta_{\rm GLX} &=& -0.05+0.90 \beta_{\rm C94}\qquad (z<0.05). \label{eqn:beta2}
\end{eqnarray}

\noindent For $\beta=-1$, the average for the sample, the difference
between the values of two slopes is 0.16 in the case of $z<0.3$,
reducing to 0.05 for $z<0.05$, in the sense that $\beta_{\rm GLX}$ is
higher. Redshift dependence exists because $\beta_{\rm C94}$ is
measured on rest-frame spectrum whereas $\beta_{\rm GLX}$ in
low-redshift studies is often determined from observed-frame UV color,
which will differ from the rest-frame UV color due to the Ly$\alpha$
contribution (e.g., \citealt{shim13}), and the fact that the UV
spectrum is not a perfect power law. The dispersion around the
relations is 0.20 and 0.16, respectively, and is driven by galaxies
having a range of UV bump strengths, coupled with the fact that the UV
bump affects $\beta_{\rm GLX}$ much more than it does
$\beta_{\rm C94}$. 

Some high-redshift studies derive the UV slope based on the FUV region
alone (1200-1800 \AA). Such $\beta$ will typically be higher
(shallower) than the 10-window slope we use here, the difference being
$\sim0.3$ at $\beta\sim -1$ \citep{calzetti01,reddy18}, but note that the
difference is not constant---it increases with $\beta$. A similar
significant difference also exists between the slope obtained from a
single continuous window covering $1268\leq \lambda \leq 2580$\AA range
($\beta_{\rm cont}$, e.g., \citealt{mclure18}) as opposed to the slope
that uses 10 windows in this same range. Using our sample we derive:

\begin{eqnarray}
\beta_{\rm cont} &= &0.55+1.21 \beta_{\rm C94}, \label{eqn:beta3}\\
\beta_{\rm C94} &=& -0.45+0.82 \beta_{\rm cont}. \label{eqn:beta4}
\end{eqnarray}

\noindent Now the difference at  $\beta\sim -1$ is 0.34. This
underscores that care must be exercised in comparing the results of
studies that measure the UV slope in different ways \citep{reddy18}.

\begin{figure}
\epsscale{1.1} \plotone{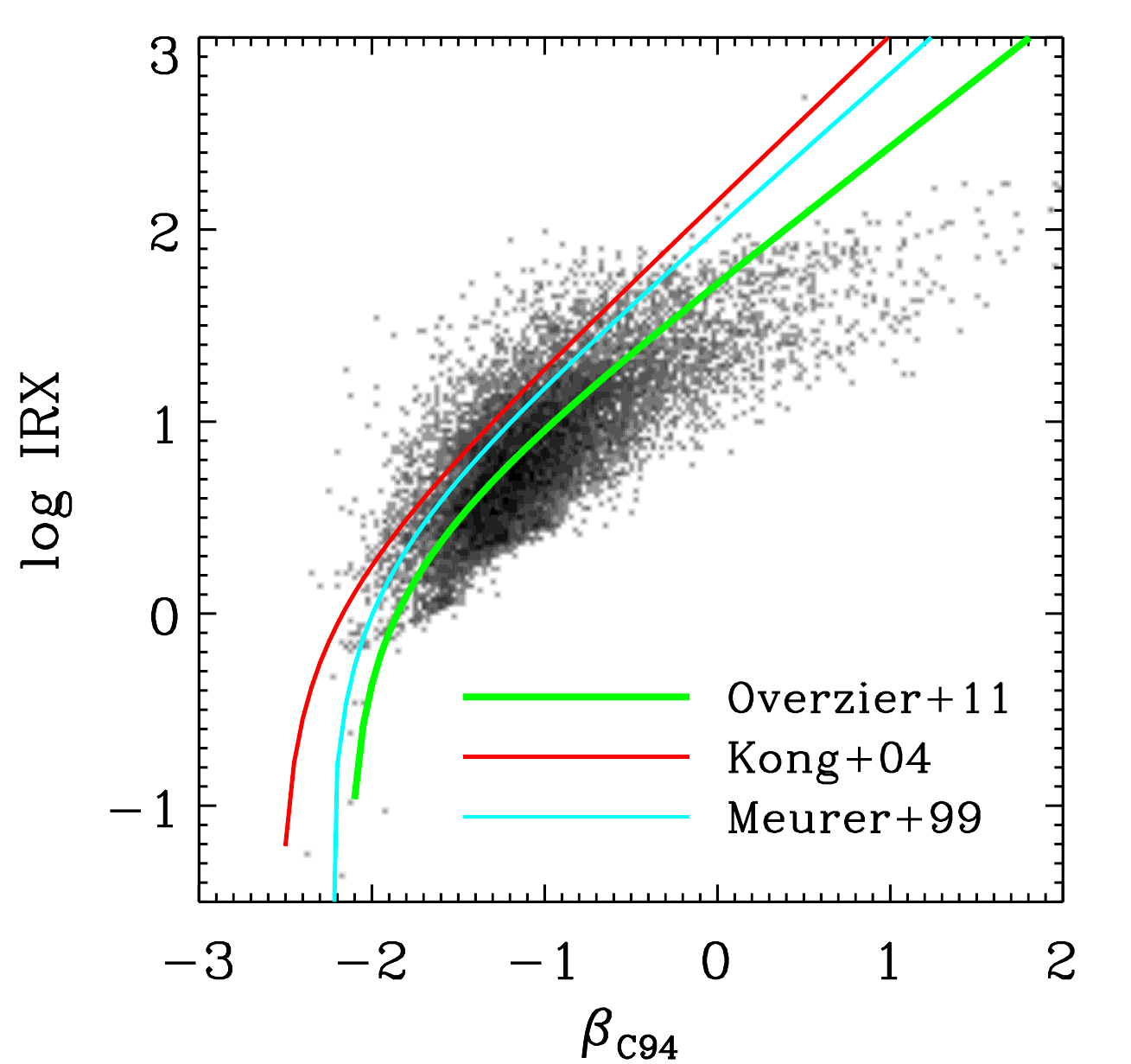}
\caption{IRX-$\beta$ data points and literature relations. Data points
  from our star-forming sample are shown with three literature
  relations derived from essentially the same sample consisting of local
  starbursts. The far-IR to total IR correction for \citet{meurer99}
  relation is included.  \citet{kong04} and \citet{meurer99} relations
  were based on {\it IUE} UV spectra taken through an aperture that missed
  $\sim 1/2$ of the total FUV flux, leading to a vertical shift with
  respect to the \citet{overzier11} starburst relation that was based
  on the same sample but using total \galex\
  photometry. \citet{overzier11} relation is used as a fiducial
  relation in all other figures in this paper.
  \label{fig:lit}}
\end{figure}

For the purposes of describing how different sample cuts lead to
different distribution of datapoints in the IRX-$\beta$ plane, we will
be introducing a fiducial IRX-$\beta$ relation. Figure \ref{fig:lit}
displays three previously published relations, all principally based
on the local starburst sample of \citet{meurer99}. The
\citet{meurer99} relation is based on 57 starburst galaxies whose UV
spectra were observed by {\it IUE}. They use the \citet{calzetti94}
methodology to measure UV slopes, but because not all spectra had the
near-UV region observed (corresponding to the tenth window),
\citet{meurer99} determined the UV slopes of {\it all} galaxies using
9 far-UV windows and then adjusting the result to what would have been
determined with all windows using a constant offset. Importantly, the
IRX in \citet{meurer99} was defined using the far-IR luminosity
(40--120 $\mu$m), whereas most consequent studies define IRX based on
the total IR luminosity (1--1000 $\mu$m). The correction factor
between them is 1.75 (\citealt{calzetti00}, rather than 1.4 reported
in \citealt{meurer99}, the difference coming from adding the sub-mm
tail to far-IR bandpass, i.e., 40--1000 $\mu$m), and has been applied
throughout. \citet{meurer99} relation adjusted to total IR luminosity
is plotted in Figure \ref{fig:lit}, and it appears not to coincide
with the mean locus of our sample, being too high (or too red).

\citet{kong04} rederive the IRX-$\beta$ relationship using essentially
the same sample as \citet{meurer99}, so it is also away from the main
locus of our data points (Figure \ref{fig:lit}). Note that
\citet{kong04} provide their relation in terms of $\beta_{\rm GLX}$,
which we convert to $\beta_{\rm C94}$ using the relation from their
Fig.\ A1, which yields a similar correction as our Eq.\
\ref{eqn:beta1}. \citet{kong04} and \citet{meurer99} relations differ
mostly in terms of the $\beta_{\rm min}$ asymptote. There are very few
points in these samples with log IRX$<0$ to allow $\beta_{\rm min}$ to
be well established. Our more extensive data suggest
$\beta_{\rm min}\approx-2.2$, a value closer to the minimum slope in
the \citet{meurer99} relation. Studies that have performed detailed UV
photometry of nearby galaxies, which includes many low-metallicity,
nearly dust-free dwarfs \citep{gildepaz07,dale09}, never find galaxies
with colors bluer than FUV$-$NUV$=-0.05$, which corresponds to
$\beta_{\rm GLX}=-2.1$, or $\beta_{\rm C94}=-2.3$. Note that our
models include SEDs with intrinsic UV slopes
$\beta_{0,{\rm min}}=-2.56$ (corresponding to 1/20 $Z_{\odot}$ and
young ($<100$ Myr) burst), but only 0.2\% of star-forming galaxies are
fit by models with $\beta_{0,\rm C94}<-2.5$.

In both the \citet{kong04} and \citet{meurer99} analyses, the FUV flux
comes from {\it IUE} measurements in $10''\times 20''$ apertures,
whereas the IR luminosity comes from {\it IRAS} measurements that
encompass entire galaxies. This inconsistence was not considered
critical by \citet{meurer99} because it was assumed that most of the
total UV flux was confined to $10''\times 20''$ aperture because the
starburst are typically compact and galaxies with large optical
diameters (but only if greater than $240''$) were explicitly excluded
from the sample. However, \citet{overzier11} measured the UV fluxes of
the large fraction of \citet{meurer99} sample using integrated
photometry from \galex, and found that the {\it IUE} measurements
missed $\sim1/2$ of the UV flux. Similar analysis and results were
obtained by \citet{takeuchi12}, and by \citet{casey14}. The latter
study performed new measurements of both starbursts and normal
galaxies. The IRX-$\beta$ relation re-derived by \citet{overzier11} is
shown in Figure \ref{fig:lit} (where their rest-frame
$\beta_{\rm GLX}$ was converted to $\beta_{\rm C94}$ using Eq.\
\ref{eqn:beta2}). The bottomline is that \citet{meurer99} and
\citet{kong04} relation should not be used with modern total UV flux
measurements, but rather their updated versions.  Our dataset is on
average well described by the \citet{overzier11} relation, and we
adopt it as our fiducial relation. The agreement is present despite
the fact that our sample is mostly not composed of starbursts, which
we will discuss in Section \ref{ssec:age}.

\begin{figure*}
\epsscale{1.1} \plotone{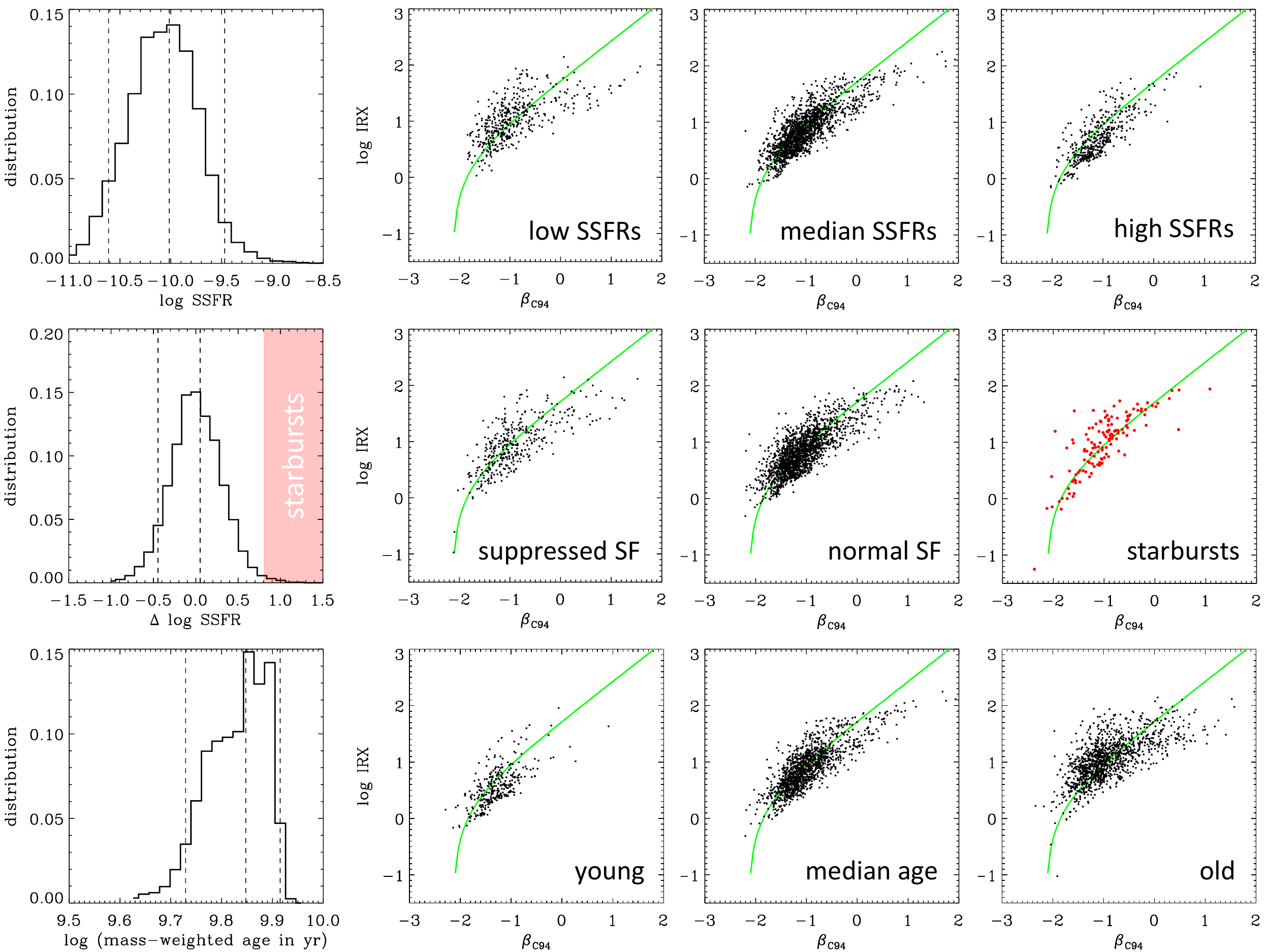}
\caption{Age-related parameters and their role in driving the
  IRX-$\beta$ scatter in star-forming galaxies. For each parameter
  (specific SFR, sSFR relative to the star-forming sequence, and
  mass-weighted population age) the distribution is shown along with
  the positions of the median, 5th and 95th percentiles (dashed
  lines). Galaxies with values within one histogram bin of these
  percentiles are plotted in separate IRX-$\beta$ panels (with the
  exception of starbursts which cover the entire shaded
  area). Selecting by these parameters does not reduce the scatter in
  IRX-$\beta$ panels, nor does it produce a notable systematic
  trend. Most importantly, both the normal star-forming galaxies and
  the starbursts are similarly well described with the
  \citet{overzier11} relation (green curve) originally derived based
  on starbursts alone. \label{fig:age}}
\end{figure*}

\begin{figure*}
\epsscale{1.1} \plotone{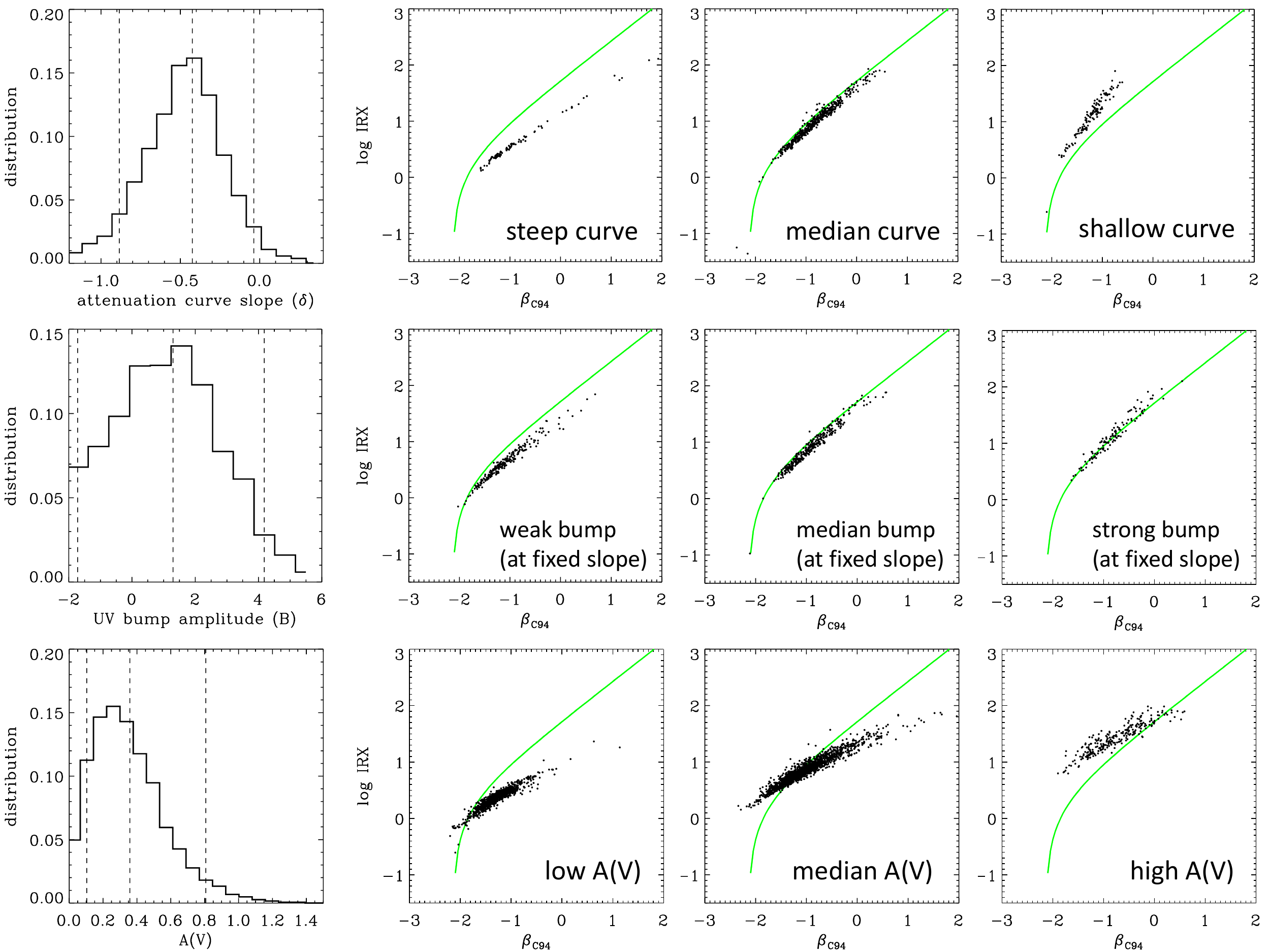}
\caption{Dust-related parameters and their role in driving the
  IRX-$\beta$ scatter in star-forming galaxies. Distributions of
  attenuation curve slopes (for a fixed UV bump), the UV bump
  amplitude (for a fixed slope) and optical attenuations is
  shown. Galaxies with values within one histogram bin of the 5th,
  50th and 95th percentiles are plotted in the the IRX-$\beta$
  panels. Attenuation curve slope is strongly correlated with the tilt
  in the IRX-$\beta$, with some additional changes in the tilt due to
  the UV bump. Attenuation curve slope itself depends on optical
  opacity \citep{pierini04,chevallard13,seon16,s18}, which is why
  selecting by $A_V$ also produces relatively tight distribution of
  points in the IRX-$\beta$ plots. The green curve represents the
  \citet{overzier11} relation. \label{fig:dust}}
\end{figure*}

\begin{figure*}
\epsscale{1.1} \plotone{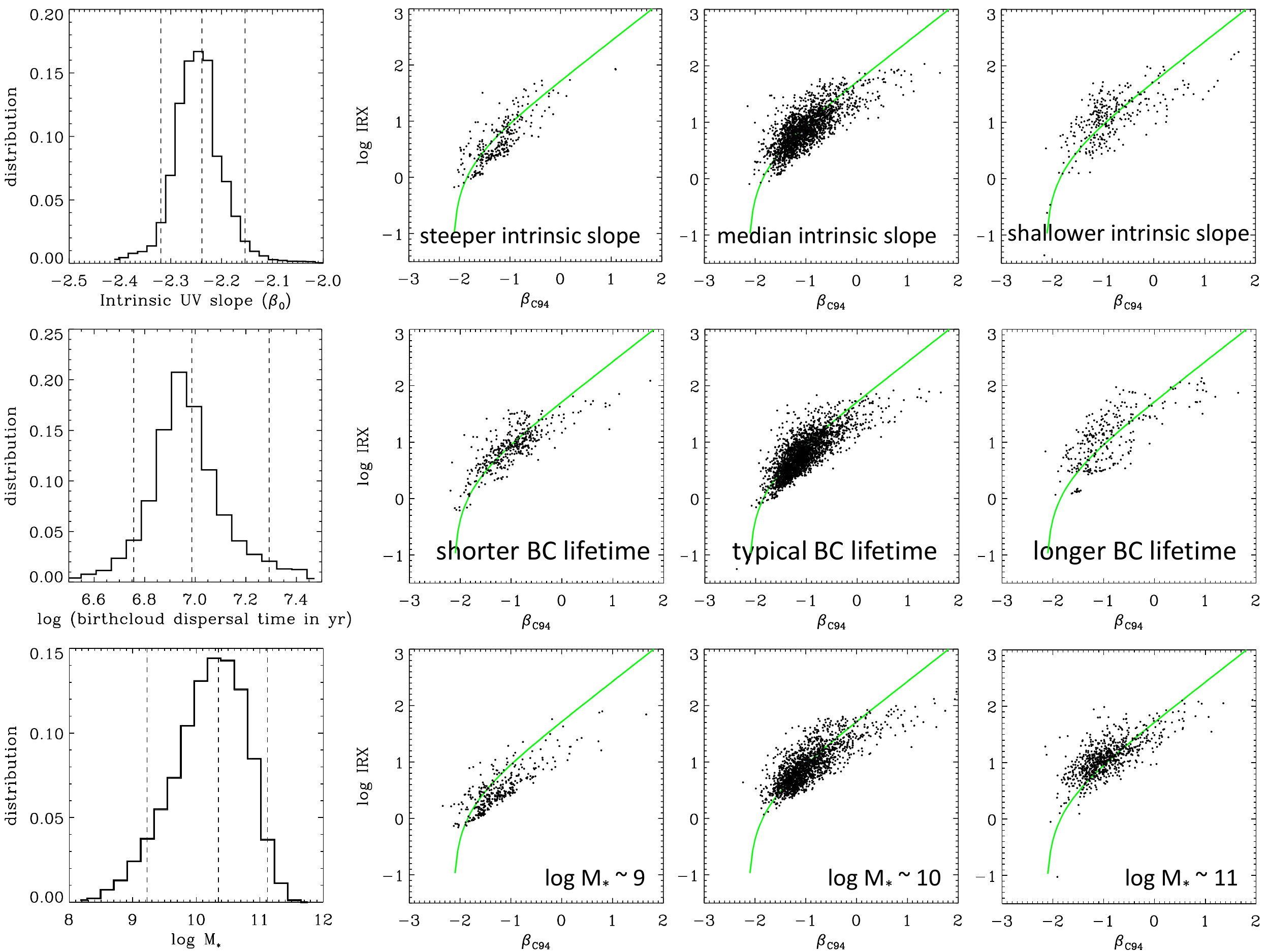}
\caption{The role of the intrinsic UV slope, birth cloud dispersal
  time and the stellar mass in driving the IRX-$\beta$ scatter in
  star-forming galaxies. The stellar mass appears to be a moderately
  strong driver of the IRX-$\beta$ distribution, which we later show
  to be because of its correlation with $A_V$. The green curve
  represents the \citet{overzier11} relation. \label{fig:other}}
\end{figure*}
green

\section{Results} \label{sec:results}

In this section we search for parameters that may govern the scatter
in the IRX-$\beta$ plane. The search is first performed in a visual
manner, and is parameterized afterwards. The visual method consists of
plotting, on IRX-$\beta$ diagrams, the slices of data having low,
median and high values of some parameter, and looking at the tightness
and the shifts/trends in distribution of points. To determine what
constitutes low/median/high values for, e.g., log sSFR, we construct
the distribution of that parameter and find values that correspond to
the 5th percentile, the median, and the 95th percentile of the
distribution. For log sSFR, the corresponding low/median/high values
are $-10.61$, $-10.02$ and $-9.47$ in units of
$M_{\odot}{\rm yr}^{-1}$. Slices have a width that we take to be 1/9
of the 90 percentile range of the distribution. In the case of sSFR,
each slice is 0.13 dex wide, so, for example, the median slice covers
$-10.08<\log {\rm sSFR} <-9.95$. In case of a uniform distribution
each bin would contain exactly 10\% of all data points. For a normal
distribution the median bin will contain 15\% of data points, whereas
the low/high bins will contain approximately 4\%.

A parameter that affects the scatter in the IRX-$\beta$ plane would
yield a much narrower distribution of points when individual slices
are plotted, compared to the full distribution of points in the
IRX-$\beta$ plane, plus the locus of points would shift considerably
when going from the low percentile to the high percentile slice. The
analysis in subsequent sections pertains to the star-forming
sample. We separately discuss passive galaxies in Section
\ref{ssec:passive}.

\subsection{Galaxy parameters related to star formation and population age} \label{ssec:age}

We start the analysis by exploring three parameters that are related
to stellar population age (Figure \ref{fig:age}), following the
original hypothesis by \citet{kong04}. The upper left panel of Figure
\ref{fig:age} shows the distribution of the specific SFRs in the
sample, spanning some 2 dex. SFRs represent averages over 100 Myr. The
distribution has a median and a peak at log sSFR$\sim -10$, with 5th
and 95th percentiles some 0.5 dex below and above it. The three 
panels to the right show where the points selected to have low, median and high
sSFR lie on the IRX-$\beta$ plane. Selecting the data by sSFR does not
reduce the scatter in the relation appreciably, and there is only a
slight shift downwards in the locus of points.

Considering that the star-forming sequence of sSFR vs.\ stellar mass
is tilted (e.g., \citealt{s07}), galaxies with higher sSFR will also
tend to be less massive, and vice versa. Thus, a better way to
distinguish galaxies by their level of activity is to look at the
relative sSFRs, i.e., the offset of sSFR with respect to the mean
sSFR-$M_*$ relation. A robust linear fit to the relation using the
current star-forming sample gives:

\begin{eqnarray}
\log {\rm sSFR} &= &-0.31 \log M_*-6.89.
\end{eqnarray}

\noindent Selecting galaxies around 5th and 50th percentile by
relative sSFR, i.e., galaxies with suppressed and with normal SF still
does not produce any tightening in the IRX-$\beta$ distribution, nor
systematic shifts of points (Figure \ref{fig:age}, middle row). The
95th percentile selected galaxies lie 0.5 dex above the main sequence,
which is not as high as the {\it IUE} starbursts of \citet{calzetti94}
and \citet{meurer99}, which lie between 0.5 and 1.5 dex above the main
sequence (based on the analysis performed for
\citealt{s18}). Therefore, in this one case we replace the 95
percentile slice with the selection that includes all galaxies with
$\Delta$log sSFR$>0.8$ dex. There are 125 such galaxies in our sample
(1\% of all star formers) and their average offset from the main
sequence is 1.0 dex, same as for the {\it IUE} starbursts. Their
IRX-$\beta$ distribution is shown in the right panel of the middle row
of Figure \ref{fig:age}. We do not see that starbursts have an offset
in the IRX-$\beta$ distribution with respect to normal or even
quiescent star-formers, though the scatter is somewhat smaller, i
agreement with the original {\it IUE} results of a relatively tight
relation. Altogether, neither the sSFR nor the relative sSFR drive the
scatter. The similarity between how normal and starbursting galaxies
are distributed in the IRX-$\beta$ plane explains why the IRX-$\beta$
relations derived from local starburst samples (but using total
\galex\ photometry) describe the average locus of general population
of star-forming galaxies so well (Fig.\ \ref{fig:lit}).

\citet{kong04}, based on their modeling, proposed that galaxies occupy
different locations in the IRX-$\beta$ plane according to their
birthrate parameter ($b$), i.e., that $b$ governs the
scatter. Birthrate parameter is defined as the ratio of the current
SFR to the past-averaged one. The latter is the ratio of the total
stellar mass produced over galaxy's history (the integral of SFR)
divided by the total age of the galaxy. However, the total age of a
galaxy, i.e., the time since the commencement of any SF in a galaxy,
is not a particularly meaningful or even measurable quantity, and one
may as well assume that all galaxies have the same total age. In that
case, the birthrate becomes equivalent to sSFR. Indeed, birthrate
parameter has been replaced in recent literature by sSFR, which we
have shown not to control the scatter.

Finally, we explore the role of the stellar population age, determined
here as the stellar-mass weighted age, and spanning the range in our
star-forming sample from 4.5 to 8.5 Gyr, with the peak towards older
systems (lower left panel of Figure \ref{fig:age}). Selecting
subsamples with younger, median and older ages and showing their
IRX-$\beta$ distributions reveals a slight trend whereby older
galaxies tend to shift upwards, but again the large scatter
remains. In the most favorable case, the stellar population age may
only weekly affect the IRX-$\beta$ distribution.

\subsection{Galaxy parameters related to dust attenuation} \label{ssec:dust}

In Figure \ref{fig:dust} we explore the role of dust attenuation
parameters on IRX-$\beta$ (still for the star-forming sample),
starting with the slope of the dust attenuation curve, parameterized
as the power-law deviation from the Calzetti curve. While the UV slope
is calculated over wavelength ranges such that the bulk of the UV bump
is excluded, we find that the bump is nevertheless wide enough to affect the UV
slope. For this reason we explore the
role of the slope and of the bump on the IRX-$\beta$ independently
from one another, by restricting the other parameter to a small range
of values.

The upper row of panels in Figure \ref{fig:dust} shows in the first
panel the distribution of the slopes (with the bump amplitude
restricted to $1<B<3$, 1/4 of the full possible range), followed by
IRX-$\beta$ panels in which the points were selected to have steep,
typical and shallow slopes. We point out that the range of slopes is
quite broad, from shallower than the Calzetti curve to steeper than
the SMC curve \citep{s18}. Selecting the subsamples by the steepness
of the slope results in dramatic reduction in scatter, with different
attenuation curve slopes occupying distinctly different regions of the
IRX-$\beta$ plane. As a matter of fact, the scatter essentially
vanishes by selecting galaxies within even narrower ranges of slopes
(and UV bump amplitudes). Different attenuation slopes correspond to
different tilts in the IRX-$\beta$ point distribution, which we will
use as a basis of new parameterization in Section
\ref{sec:eta}. Galaxies with steeper attenuation curve slopes tend to
lie below the \citet{overzier11} relation, whereas the galaxies with
shallow slopes (similar to the Calzetti curve slope or MW curve) tend
to lie above the \citet{overzier11} relation.

In the middle row of panels in Figure \ref{fig:dust} we keep the
attenuation curve within a restricted range ($-0.5<\delta<-0.3$) and
explore any additional trends due to the varying strength of the UV
bump. The distribution in the IRX-$\beta$ plane is tight because of
the restricted range of attenuation curve slopes. Plotting the
galaxies with weak, median and strong bump reveals a moderate shift in
the locus. The bump apparently does systematically affect the
measurement of the UV slope, even when it was derived over the 10
windows. Specifically, we find that for each unit of amplitude of the
bump, the UV slope is shifted by $d \beta_{\rm C94} / d B = -0.064$.

Altogether, the conclusion is that {\it the shape of the attenuation
  curve, primarily its slope, is the principal driver of the scatter
  in the IRX-$\beta$ diagram.} Consequently, the question of what
drives the IRX-$\beta$ scatter becomes the derivative of the question
of {\it what drives the diversity of slopes of the attenuation curve.}
\citet{s18} has focused on the latter question and found that the
principal connection exists between the attenuation curve slope and
the optical opacity (e.g., $A_V$), with more transparent galaxies
having steeper slopes, presumably because the wavelength-dependent
scattering dominates over the absorption
\citep{chevallard13}. Radiative transfer modeling also shows the
connection between the attenuation curve slope and optical opacity
\citep{witt00,pierini04,seon16}. The bottom panels of Figure
\ref{fig:dust} explicitly explore how selecting galaxies by $A_V$
translates to the IRX-$\beta$ plane. The distribution of points is
relatively narrow, though not as narrow as for the attenuation slope,
and the galaxies with different $A_V$ tend to have a similar tilt
offset from one another, unlike the changing tilt when varying the
attenuation curve slope. Thus, while related, the two factors, dust
attenuation curve slope and $A_V$, are not equivalent. We shall return
to the connection between the dust attenuation curves and $A_V$ in
Section \ref{sec:disc}.

Our sample contains very few ULIRGs as they are intrinsically
rare. Local ULIRGs tend to span $1.5<\log {\rm IRX}<3$
\citep{howell10}, whereas in our sample log IRX does not exceed 2. In
light of our results, the special location of ULIRGs in the
IRX-$\beta$ plane
\citep{goldader02,buat05,reddy10,penner12,oteo13,casey14} may not
require a separate explanation. High values of IRX are due to high
opacities, but their dispersal along $\beta$ should still be the
result of the diversities of attenuation curve slopes
\citep{lofaro17}. High optical opacities of ULIRGs implies shallow
attenuation curves, which are located mostly to the left of the mean
relation, as confirmed in the actual data (e.g.,
\citealt{goldader02,howell10}).

\begin{figure*}
\epsscale{1.2} \plotone{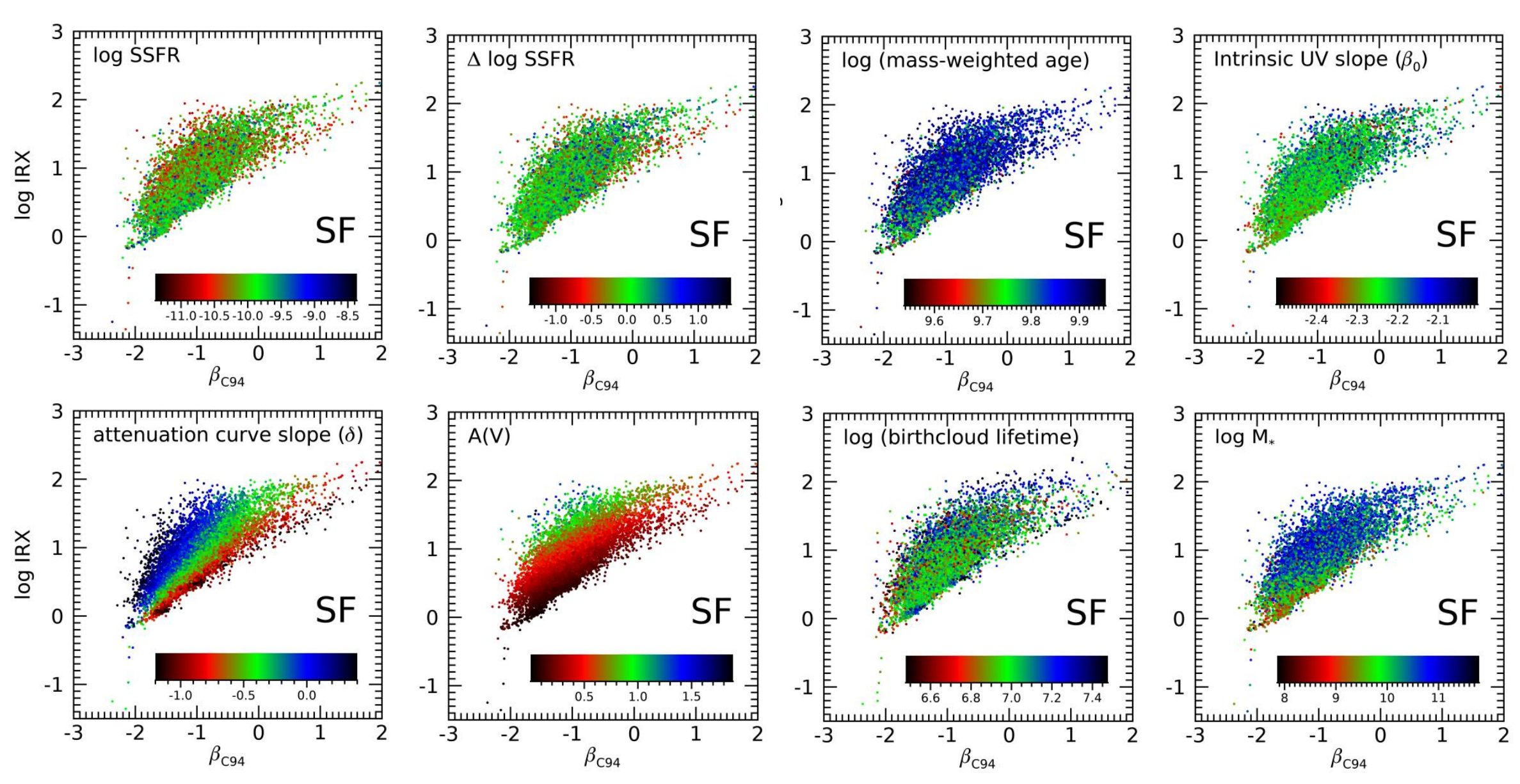}
\caption{IRX-$\beta$ color-coded by the value of a galaxy
  parameters, for the star-forming sample. The parameter is indicated
  in each panel, with appropriate colorbar legend showing the range of
  parameter value. All star-forming galaxies are shown in all
  panels. \label{fig:colorcoded_sf}}
\end{figure*}

\begin{figure}
\epsscale{1.1} \plotone{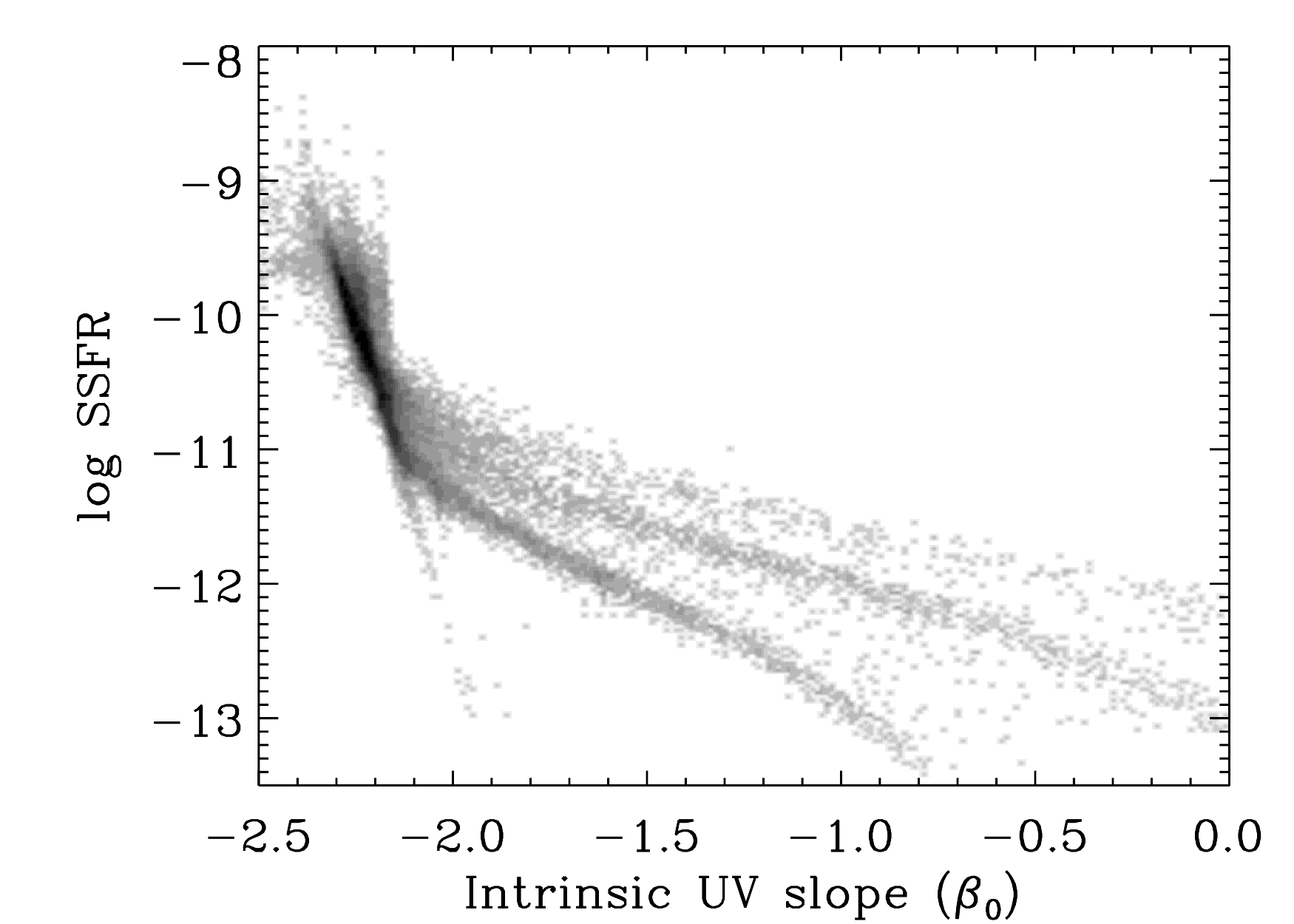}
\caption{Relationship between the specific SFR and the intrinsic UV
  slope. The sample now includes all galaxies (star-forming and
  passive). Passive (low sSFR) galaxies exhibit a large range of
  shallow UV slopes, that depend on sSFR as well as on the stellar
  metallicity (discrete tracks). \label{fig:ssfr_beta0}}
\end{figure}

\subsection{Other galaxy parameters} \label{ssec:other}

Other parameters that have been proposed to affect the IRX-$\beta$
include: (1) the intrinsic UV slope \citep{boquien12}, which is
sensitive to SF history, but, as we will see, also on the stellar
metallicity, (2) the stellar mass (e.g., \citealt{alvarez16}), (3) the
lifetime of birth clouds and (4) the fraction of dust absorbed in
birth clouds \citep{cf00}, as well as (5) the galaxy inclination
\citep{wang18}.

The intrinsic UV slope $\beta_0$ is the slope that one would observe
in the absence of dust. We determine it from the SED fitting with the
same 10-window methodology used for $\beta_{\rm C94}$. The upper left
panel of Figure \ref{fig:other} shows that the distribution of
intrinsic slopes of star-forming galaxies is rather narrow, with 5 and
95 percentile values lying within 0.1 of the median at
$\beta_0=-2.24$. Selecting the galaxies that span that range produces
only a slight shift in the locus of points, with the scatter that is
similar to the scatter of the full sample. Distribution of intrinsic
slopes extends to less steep values (up to $\beta_0\sim -1$) only when
passive galaxies (log sSFR$<-11$) are included, which will be
discussed in Section \ref{ssec:passive}.

Following \citet{cf00}, our default SED fitting assumes that birth
clouds, which suffer higher attenuation than ambient ISM, disperse
after 10 Myr. However, \citet{cf00} have showed model tracks with different
values of birth cloud dispersal time showing a shift in the
IRX-$\beta$ plane. We thus produce a separate run where we allow the
split between young and old stars (i.e., birth cloud dispersal time)
to span a range from 3 to 30 Myr. The middle left panel of Figure
\ref{fig:other} shows that the distribution is indeed highly peaked at
10 Myr, and furthermore that selecting galaxies by shorter and longer
dispersal times does not reduce the scatter nor produce trends in the
IRX-$\beta$ plane.

\begin{figure*}
\epsscale{1.2} \plotone{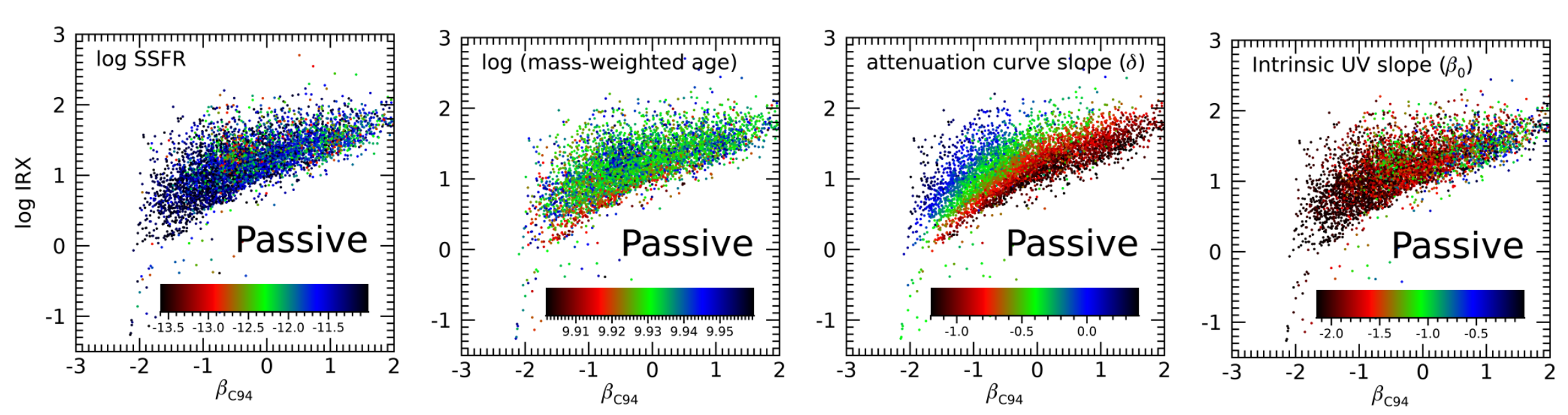}
\caption{IRX-$\beta$ diagrams of passive galaxies, color-coded by the
  value of a galaxy parameters. Passive galaxies are selected to have
  log sSFR$<-11$. The parameter is indicated in each panel with
  appropriate colorbar legend showing the range of values for the
  parameter. Again, the slope of the attenuation curve is the leading
  driver of the scatter. \label{fig:colorcoded_pass}}
\end{figure*}

\begin{figure}
\epsscale{1.1} \plotone{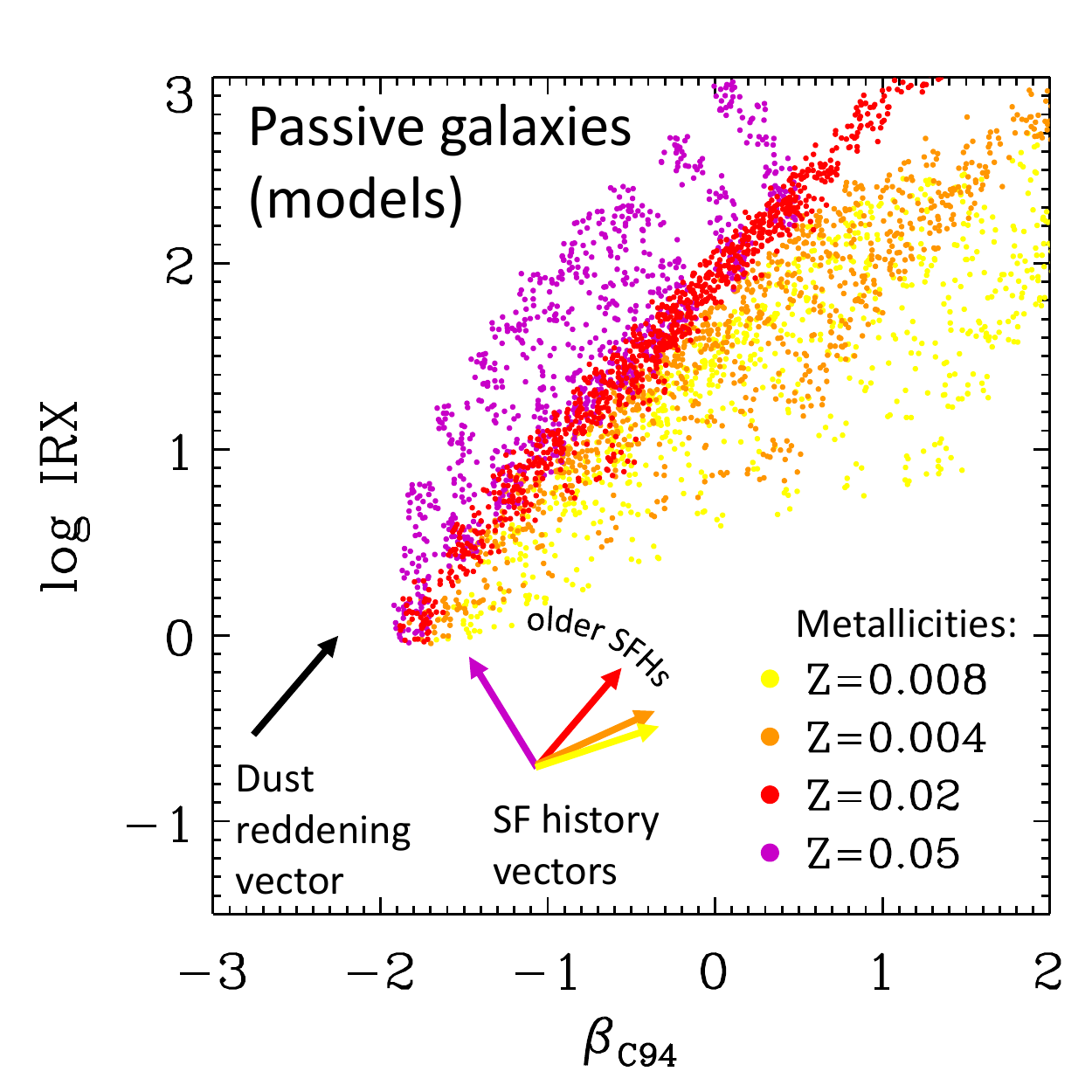}
\caption{Model IRX-$\beta$ points for passive galaxies having a fixed
  dust attenuation curve. Even though the dust attenuation curve is
  fixed, ($\delta=-0.4$, $B=1$). Passive galaxies (log sSFR$<-11$) can
  exhibit large dispersion in the IRX-$\beta$ plane, due to a large,
  metallicity-dependent spread of intrinsic UV slopes corresponding to
  old SF histories., Fortuitously, the star-formation history vector
  for galaxies with the solar stellar metallicity ($Z=0.02$), which is
  the bulk of the sample, are well aligned with the dust attenuation
  vector, resulting in little scatter due to the spread of intrinsic
  UV slopes.  \label{fig:model_pass}}
\end{figure}

Similarly, in our nominal SED run we keep the ratio of the $V$-band
attenuation affecting older stars (those which left the birth clouds)
to be some fixed fraction of the attenuation affecting young stars.
Again, the modeling of \citet{cf00} suggested that allowing this
fraction to vary may lead to trends in the IRX-$\beta$ plane, so we
produce a run where the ratio is not fixed. Selecting the galaxies
with low ratio necessarily selects galaxies with low $A_V$, which
restricts them to low IRX values. Otherwise, there is no clear trend
for the different values of the old to young attenuation ratio, and we
do not present the associated plots.

We also examined (but do not present) the IRX-$\beta$ plots for
galaxies selected to have different inclinations, measured as $b/a$
axis ratios. Highly inclined edge-on galaxies ($0.2<b/a<0.3$) do tend
to lie in the IRX-$\beta$ diagram somewhat above the face-on galaxies
($b/a>0.9$), and to exhibit a larger scatter in IRX. Similar trends
have been recently seen at $z\sim1.5$ by \citet{wang18}. \citet{s18}
have shown that edge-on galaxies tend to have somewhat shallower
attenuation curves than the face-on galaxies, and that the difference
in attenuation curve slopes is entirely due to the higher average dust
content, i.e., greater $A_V$, of edge-on galaxies. Once $A_V$ is
accounted for by fixing it, the dependence goes away.

Finally, we look at the IRX-$\beta$ dependence on the stellar mass
(bottom row of Figure \ref{fig:other}. Selecting the slices of data
around the 5th, 50th and 95th percentile roughly corresponds to
galaxies with masses around $\log M_*=9$, 10 and 11. While the
substantial scatter in these mass-selected slices remains, the shift
in the locus is more pronounced than for any of the non-dust
parameters that we have considered. The mass dependence of similar
magnitude was previously reported in a $z\sim3$ sample, from a
stacking analysis of \citet{alvarez16}, which motivated them to
suggest that the IRX$-M*$ relation should be combined with
IRX-$\beta$. The dependence of the slope of the attenuation curves on
$M_*$ was discussed in \citet{s18}, and, as in the case of
inclination, it was shown that this dependence is entirely due to the
more-massive galaxies having higher $A_V$, a parameter that most
strongly drives the slopes of the attenuation curves. We revisit this
analysis in the context of IRX-$\beta$ in Section \ref{sec:eta}.

An alternative way of showing the relationship between IRX-$\beta$ and
the various parameters is to color-code the datapoints according to
the value of the parameter (Figure \ref{fig:colorcoded_sf}). The
strongest trend is with respect to the slope of the attenuation curve,
followed by the trend with respect to $A_V$. Other trends exist as
well (most notably with respect to the stellar mass), but are small
compared to the galaxy-to-galaxy scatter.

\subsection{Passive galaxies} \label{ssec:passive}

In Figure \ref{fig:other} we have shown that the range of intrinsic UV
slopes ($\beta_0$) is quite small ($\sim 0.2$) and therefore cannot
have a significant role in driving the scatter in the IRX-$\beta$
relation. This is because in our analysis so far we have focused on
star-forming galaxies, the type of galaxies for which the IRX-$\beta$
is expected to be useful. Our full sample, however, does include
passive galaxies ($-13<\log {\rm sSFR}<-11$), with the data needed to
place them on the IRX-$\beta$ plot. We first show, in Figure
\ref{fig:ssfr_beta0}, the relationship between the specific SFR and
the intrinsic UV slope for the full sample (star forming and passive),
which confirms that actively star-forming galaxies have a small range
of $\beta_0$. The range becomes much wider (towards shallower values
of $\beta_0$) for passive galaxies, extending all the way to
$\beta_0\approx 0$. In addition to the sSFR, the intrinsic UV slope of
passive galaxies depends very strongly on the stellar metallicity,
with metal-poor galaxies having shallower intrinsic UV slopes (higher
$\beta_0$).

In Figure \ref{fig:colorcoded_pass} we use color-coded values to look
at the role of several select parameters in driving the IRX-$\beta$
scatter of 5180 passive galaxies. The galaxies are selected to have
log sSFR$<-11$, regardless of their BPT diagram classification First,
we notice that the overall distribution of the points is not too
dissimilar to that of the actively star-forming galaxies. Next, it is
again the slope of the attenuation curve that presents the strongest
trend. Stratification according to $\beta_0$ is present only in the
sense that the observed slope must be shallower than the intrinsic
slope ($\beta_{\rm C94}>\beta_0$), but for any given intrinsic slope
the distribution of points to the right of that value is quite broad.

This empirical result that even passive galaxies strongly segregate in
the IRX-$\beta$ plane by the slope of the attenuation curve is
surprising, given that the wide range of intrinsic UV slopes should
largely smear the correlation. To understand the root causes behind
this result we turn to stellar population models. In Figure
\ref{fig:model_pass}, we show model IRX-$\beta$ values for galaxies
spanning a range of passive star-formation histories (log sSFR$<-11$)
and having various attenuations, but assuming a fixed dust attenuation
curve ($\delta=-0.4$, $B=1$). As expected, the points disperse widely,
driven by a large range in $\beta_0$. The points do not just shift
along $\beta$ for older galaxies, but also move up, because the IRX
will include the increased contribution from the dust heating of older
stars. Furthermore, the direction in which the points corresponding to
shallower $\beta_0$ disperse is highly dependent on the stellar
metallicity. For super-solar metallicity ($Z=0.05$) the older galaxies
even get bluer in the UV, perhaps because of the metal line blanketing
of turn-off stars in the NUV. The great majority of passive galaxies
in our sample have near-solar metallicity ($Z=0.02$). For solar
metallicity the direction of the spread due to stellar population
being older (red arrow) matches almost exactly the dust attenuation
vector. This effectively means that the wide range of $\beta_0$ will
not lead to an additional scatter. This fortuitous alignment will
preserve the tight correlation between the IRX-$\beta$ point
distribution and the slope of the dust attenuation curve.

It should be pointed out that for passive galaxies the relationship
between $A_{\rm FUV}$ and IRX becomes strongly dependent on the
population age, because of the dust heating from old stars
\citep{cortese08}, so the derivation of $A_{\rm FUV}$ becomes
uncertain from IRX alone, and it is better to base the SFR corrections
on the methods that do not involve IR emission or use full SED
fitting.

\begin{figure}
\epsscale{1.1} \plotone{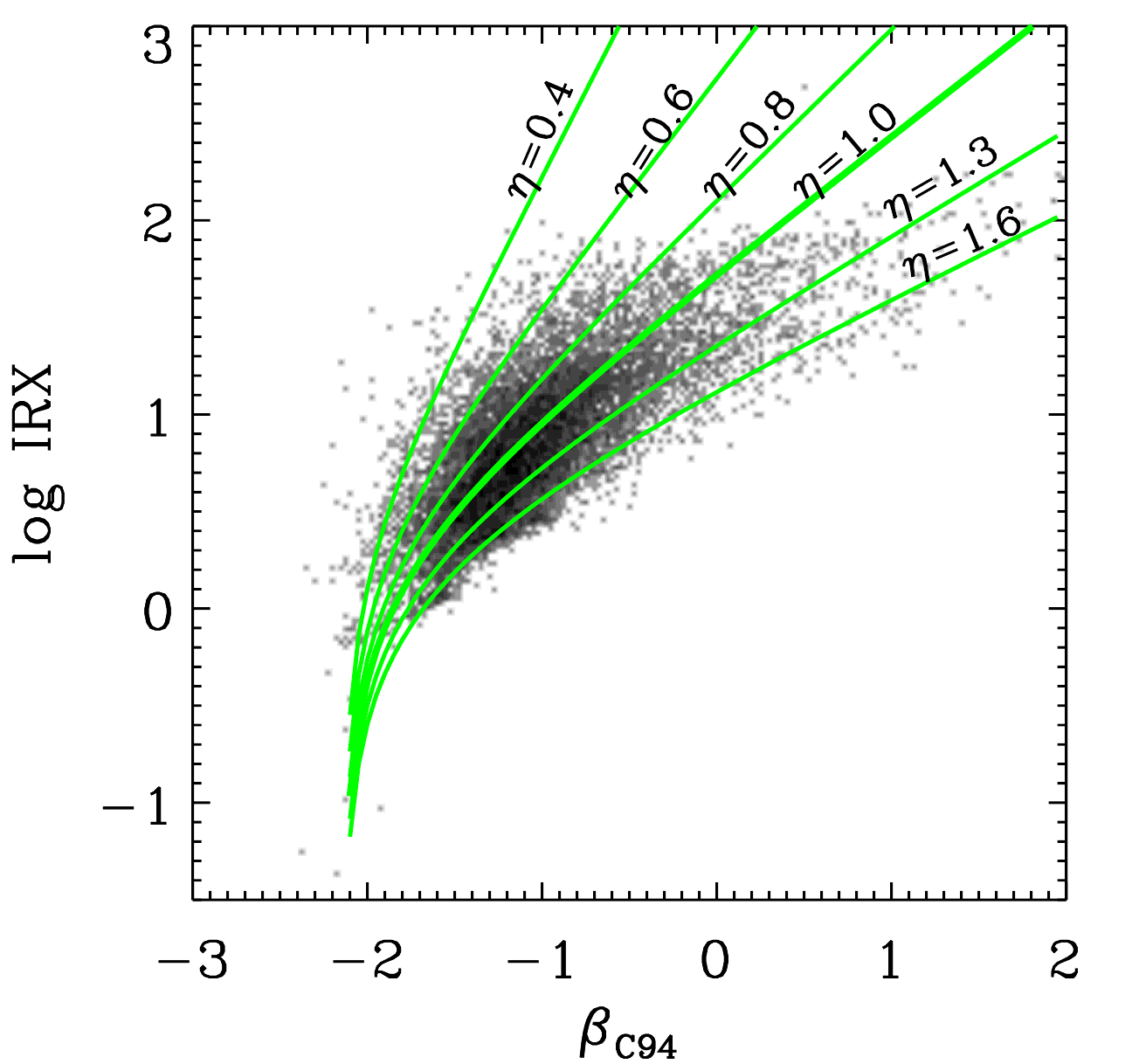}
\caption{Parameterization of IRX-$\beta$ scatter useful for measuring
  the attenuation curve slope. The parameterization consists in
  modifying the tilt of the \citet{overzier11} relation ($\eta=1$),
  using parameter $\eta$. The spread of the sample is well described
  by a family of modified curves. \label{fig:scheme}}
\end{figure}

\begin{figure*}
\epsscale{1.0} \plotone{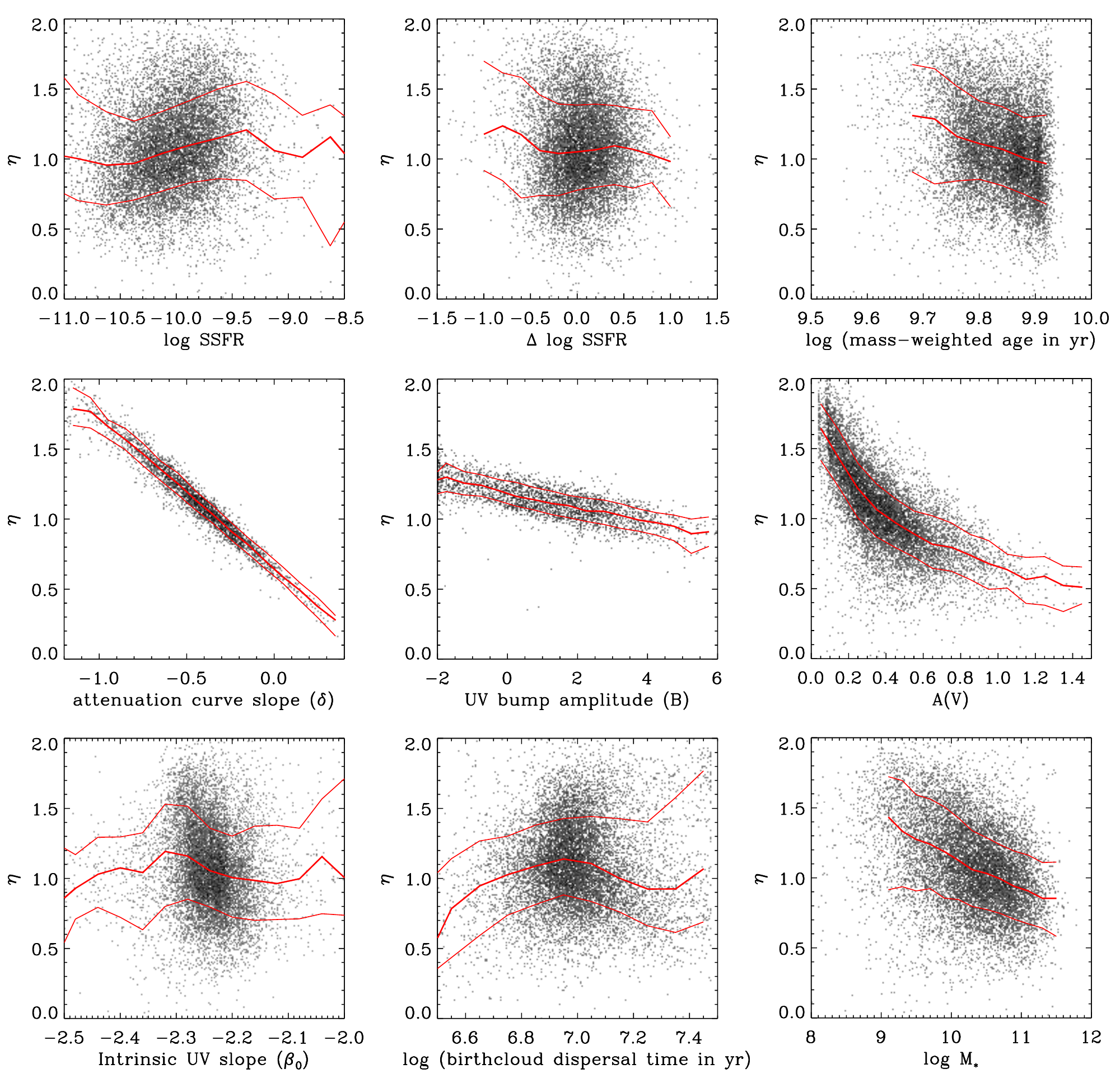}
\caption{The tilt (scatter) with respect to the fiducial IRX-$\beta$
  relation vs.\ galaxy parameters (for star-forming sample). Parameter
  $\eta$ is tightly correlated with the slope of the dust attenuation
  curve $\delta$, allowing it to be determined directly. The strongest
  other correlation is with $A_V$, which is the main driver of the
  diversity of attenuation curve slopes.. Note that in order to
  separate the effect of the slope and of the bump, each is kept fixed
  (i.e., within a restricted range) when the other is allowed to
  vary. Weaker trends seen with respect to other parameters (e.g.,
  stellar population age and stellar mass) are not primary trends (see
  Fig.\ \ref{fig:eta_avfixed}). Red lines show the running median and
  the 16th and the 84th percentiles. \label{fig:eta}}
\end{figure*}

\section{New IRX-$\beta$ parameterization} \label{sec:eta}

Informed by the results presented in the previous section, whereby the
scatter in the IRX-$\beta$ plane essentially vanishes by fixing the
dust attenuation curve, in this section we introduce new
parameterization of IRX-$\beta$ values, to serve two purposes:
allowing the slope of the attenuation curve to be measured directly
from IRX and $\beta$ values, and allowing the investigation of more
subtle trends in the IRX-$\beta$ scatter and, consequently, the dust
attenuation curve.

Looking at the upper panels of Figure \ref{fig:dust}, we see that
different attenuation curve slopes select objects having a different
tilt with respect to the \citet{overzier11} relation. The change of
the tilt of an IRX-$\beta$ relation, without the change of
$\beta_{\rm min}$, can be accomplished by scaling its
exponent. Specifically, we scale the exponent of \citet{overzier11}
IRX-$\beta$ relation ($0.78\beta_{\rm GLX}+1.54 = 0.4A_{\rm FUV}$) by
$1/\eta$.  Figure \ref{fig:scheme} shows a family of curves produced
by different $\eta$ parameters. Points above (steeper than) the
\citet{overzier11} relation have $\eta<1$. Most data points lie within
$0.4<\eta<1.6$ range. Thus, we can describe the entire IRX-$\beta$
distribution using one parameter that informs us how much each galaxy
deviates from the fiducial IRX-$\beta$ relation.

The value of parameter $\eta$ of any IRX-$\beta$ data point will be:

\begin{eqnarray}
\eta &=& \frac{\log \left(\frac{\rm IRX}{1.70}+1\right)}{0.70\beta_{\rm C94}+1.50},
\end{eqnarray}

\noindent where $\beta_{\rm GLX}$ featured in \citet{overzier11} was converted to
$\beta_{\rm C94}$ using Equation \ref{eqn:beta1}. 

Figure \ref{fig:eta} shows the dependence of $\eta$ on different
galaxy parameters discussed in Section \ref{sec:results}. As expected,
and as desired, the strongest trend of the tilt $\eta$, with the least
amount of scatter, is with respect to the dust attenuation curve
$\delta$. The tight correlation means that the slope of the
attenuation curve (expressed as its departure from the curve of the
Calzetti slope) can be determined directly from IRX-$\beta$ using a
simple linear relation:

\begin{eqnarray}
\delta &= &0.79 -1.05\eta, \label{eqn:deltaeta}
\end{eqnarray}

\noindent where the coefficients were determined based on a robust
bisector fit, and without placing any restrictions on the UV bump
amplitude. The scatter around the relation is 0.12, driven almost
entirely by the effect of varying UV bump amplitude on
$\beta_{\rm C94}$. The scatter is restricted to this value also
because of the correlation between the slope and the UV bump amplitude
\citep{kriek13,s18}. We underscore the fact that the slope $\delta$
corresponds to intrinsic curves in age-dependent application of the
modified Calzetti curve. The slope of an {\it effective} attenuation
curve will be $\delta_{\rm eff} = \delta-0.20$.

\begin{figure*}
\epsscale{1.0} \plotone{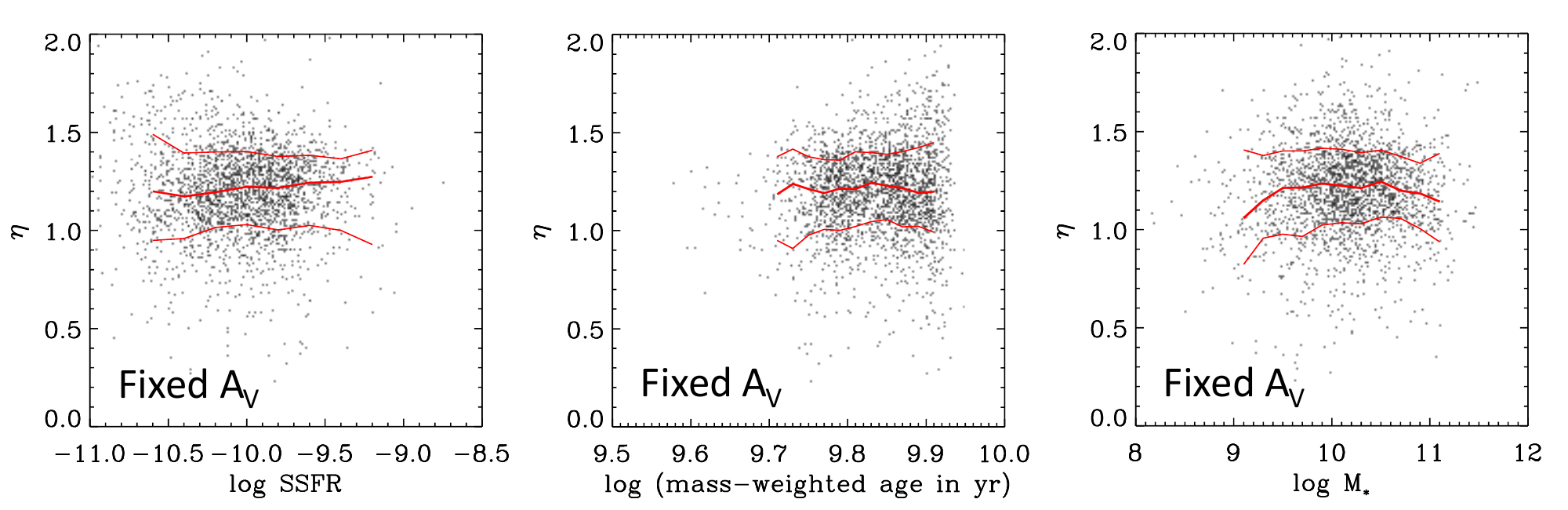}
\caption{The tilt (scatter) of IRX-$\beta$ vs.\ select galaxy
  parameters (for star-forming sample) The trends of $\eta$ vs.\ sSFR,
  the stellar population age, and the stellar mass, seen in Fig.\
  \ref{fig:eta}, disappear once the optical attenuation $A_V$ is fixed
  (here we use $0.2<A_V<0.3$). This shows that these parameters have
  an effect on the IRX-$\beta$ plot (and dust attenuation curves) only
  through their correlation with $A_V$, and do not represent
  independent factors driving their scatter. Red lines show the
  running median and the 16th and 84th
  percentiles. \label{fig:eta_avfixed}}
\end{figure*}

Figure \ref{fig:eta} also allows us to investigate, in more detail
than it was possible from Figures 2-5, any subtle effects that
different parameters may have on the locus and the scatter of data
points in the IRX-$beta$ plane. The advantage of the new
parameterization compared to the parameterization using the
perpendicular distance from a fiducial relation \citep{kong04}, is
that the tilt directly informs us about the trends in the dust
attenuation curve. The strong trend of $\eta$ vs.\ $A_V$ was already
discussed in the context of the optical opacity being a strong driver
of the dust attenuation curve, and will be discussed more in Section
\ref{sec:disc}. There is a weaker trend of $\eta$ as a function of
sSFR as well as the function of the average stellar population age and
the stellar mass. However, neither of these trends are primary. They
go entirely away once $A_V$ is fixed, as shown in Figure
\ref{fig:eta_avfixed}, where the optical opacity is restricted to
values $0.2<A_V<0.3$. Thus, the trend of $\eta$ with respect to the
stellar mass is entirely the result of the well-known underlying trend
between $M_*$ and $A_V$. Simular explanation holds for a relatively
weak trend of $\eta$ with inclination (plot not shown), which also
disappears when $A_V$ is accounted for.

We conclude that different global galaxy properties may in some cases
correspond to somewhat different attenuation curves (albeit with quite
large scatter), but what drives the variety of attenuation curves is
fundamentally due to the differences in optical opacity, which itself
depends on the amount of dust and dust geometry. This conclusion is
further born out by performing a linear regression of $\eta$ against
other parameters and finding that only $A_V$ reduces the scatter
substantially and independently from other parameters.

\section{Dust correction recipe and the impact of scatter on SFR
  determination}\label{sec:recipes}

The relatively large scatter in the IRX-$\beta$ relation may appear to
bring into question its utility for deriving dust-corrected SFRs. To
determine the impact of the scatter, we compare the SFRs and sSFRs
that would have been derived using several popular IRX-$\beta$-related
recipes and the accurate SFRs from the SED+LIR fitting. We focus on
star-forming galaxies and specifically look at the \citet{overzier11}
relation that we have used throughout the paper and two
$A_{\rm FUV}$-(FUV$-$NUV) relations: one from \citet{s07} (Eq.\ 5) and
another from \citet{hao11}. We find that these (s)SFRs have a random
scatter of between 0.14 and 0.18 dex with respect to the (s)SFRs from
the SED+LIR fitting, with little or no zero point offset, and only a
mild non-linearity (tilt). We conclude that the use of a fixed, but
nevertheless appropriate IRX-$\beta$ or $A_{\rm FUV}$-(FUV$-$NUV)
relation does not render the derived SFRs useless, just adds a random
error of some 0.15 dex. This assessment is based on our sample, which
is a mix of galaxies spanning a range of UV slopes
$-2\lesssim \beta \lesssim 1$. Beyond $z=4$, typical galaxies that one
detects in deep surveys have only blue UV slopes, $\beta \lesssim -1$
\citep{overzier08,bouwens09}, which is the combined effect of lower UV
luminosities and the lower dust content for galaxies of fixed
luminosity. The joint effect of this evolution is that the derivations
of the SFR will be less affected by the dispersion in the
IRX-$\beta$. Restricting the above exercise to $\beta<-1.5$, the added
uncertainty due to scatter becomes 0.12 dex.

Our results show that truly accurate SFR correction is only possible
if the dust attenuation curve is known. If it is not, i.e., when IR
data are not available, the IRX-$\beta$ or $A_{\rm FUV}$-$\beta$ could
potentially be expanded to include a dependence on some observable
paramater(s) that can hopefully serve as a proxy for the attenuation
curve. Of the various observable parameters that we considered, the
one with the strongest correlation with the dust attenuation curve is
the stellar mass (Figure \ref{fig:eta}), because of its correlation
with $A_V$, which in turn drives the slope of the curve. Including the
stellar mass in the dust correction recipe, such as suggested by
\citet{alvarez16}, can make the correction somewhat more accurate. We
perform a linear regression and find the following relation:

\begin{eqnarray}
A_{\rm FUV} &= &1.117\, \beta_{\rm C94}+0.262\, (\log M_*-10) +2.92. \label{eqn:recipe}
\end{eqnarray}

\noindent The scatter about the relation is 0.39 mag, somewhat better
than when the mass term is ignored (0.41 mag). The exact benefit will
depend on the mass distribution of the sample. It should be pointed
out that the $A_{\rm FUV}$-$M_*$ relation (i.e., IRX-$M_*$) alone is
considerably inferior to the $A_{\rm FUV}$-$\beta$ relation, with 0.67
mag of scatter. Another possibility to refine the estimate of SFR dust
correction is to try to use the nebular attenuation, e.g., the Balmer
decrement (BD), even though it is a rather rough proxy for the stellar
continuum attenuation (see Fig.\ 12 in \citealt{s18}). Including the
Balmer optical depth (based on SDSS spectroscopy) provides somewhat
tighter correlation than including the stellar mass:

\begin{eqnarray}
A_{\rm FUV} &= &1.006\, \beta_{\rm C94}+1.111\, \tau_{\rm Bal} +2.47, \label{eqn:recipe2}
\end{eqnarray}

\noindent where $\tau_{\rm Bal}$ is the Balmer optical depth
($\tau_{\rm Bal} = \ln ({\rm BD}/2.86)$) and the scatter around the
relation is 0.36 mag, a 20\% reduction in variance with respect to a
relation that depends on $\beta$ alone. As in the case of the stellar
mass, the Balmer optical depth alone is only poorly correlated with
$A_{\rm FUV}$, with the scatter of 0.59 mag. A relation that would
include both the stellar mass and the Balmer optical depth (in
addition to $\beta$) has almost no additional benefit because of the
mutual correlations.

The above recipes should be largely redshift-independent out to
$z\sim3$, since there does not appear to be much evolution in the
average IRX-$\beta$ relation out to that redshift \citep{alvarez16},
assuming unbiased sample selection. If SFRs are being determined from
the SED fitting, which is preferable to the above recipes because it
takes into account all of the available information, the diversity of
attenuation curves can be taken into account by applying in the SED
fitting code the mass-dependent attenuation curves provided in
\citet{s18}.

\section{Discussion} \label{sec:disc}

%
\subsection{Do normal galaxies follow a different IRX-$\beta$ relation
  from starbursts?}

\citet{kong04} originally suggested that starbursts and normal
star-formers follow different IRX-$\beta$ relations, both based on the
observations and on the IRX-$\beta$ modeling. Empirically, they have
shown that normal SF galaxies tend to be to the right of starbursts
(redder UV color at fixed IRX) and that the offset may correlate with
the age-sensitive D4000 index. We do not find any such trend, which,
in the context of our results, means that the two populations have on
average similar attenuation curves, as has also shown directly in
\citet{s18}. In \citet{kong04}, UV observations of starbursts come
entirely from {\it IUE} sample without aperture corrections, whereas
their SF sample consists of {\it IUE} observations that have been
aperture-corrected, or of observations from UV facilities with larger
apertures. Discrepant UV measurements could systematically offset the
IRX value because the IR luminosity in all cases comes from integrated
{\it IRAS} measurements. However, the possibility that the age may
lead to trends in IRX-$\beta$ has additionally been supported by
\citet{kong04} modeling, showing that the offset from the starburst
relation is correlated with the birthrate parameter (akin to sSFR, see
Section \ref{ssec:age}). In light of the results from the current
study, that it is the attenuation curve that ultimately matters, it
could be that the trend with the birthrate parameter is the
consequence of the assumed attenuation model in \citet{kong04}, the
two-component young/old populations model of \citet{cf00}, where the
attenuation slope is strongly dependent on the population
age. Dependence on age is strong in \citet{cf00} type models because
the underlying extinction curve is fixed. Once it is allowed to vary,
as is likely the case in reality, the age becomes a minor factor
\citep{popping17}. Indeed, like the current study, many other studies
that tried to find the dependence on the birthrate parameter, or other
age indicator, did not meet with success (e.g.,
\citealt{burgarella05,seibert05,panuzzo07,johnson07}).

Despite not finding the dependence on the birthrate parameter {\it
  within} their sample, \cite{panuzzo07} nevertheless show a large
offset between their entire sample (with UV measurements from \galex)
and the \citet{meurer99} starburst {\it relation} derived from {\it
  IUE}-observed starbursts. Similar offset between normal star formers
and the line describing starbursts, was presented in many other
studies \citep{buat05,seibert05,gildepaz07,takeuchi10}. At face value
these results indicate that starbursts are somehow different, despite
the lack of a trend involving the birthrate parameter (the value of
which should be much higher in starbursts than in the normal
galaxies). The resolution to this issue has been found by
\citet{overzier11} and \citet{takeuchi12} (and confirmed later by
\citealt{casey14}), who have remeasured the UV fluxes of the original
{\it IUE} starburst sample but now using \galex\ images, and have
found that the total FUV (and NUV) flux is 2-3$\times$ greater than
what has been measured in the small aperture of {\it IUE}. The flux
loss makes IRX lower, shifting the starburst relation upwards. The
possibility of a systematic aperture bias was already suggested by
\citet{cortese06}, \citet{gildepaz07} and \citet{boissier07}. The
rederived starburst relations of \citet{overzier11} and
\citet{takeuchi12} are shifted downward with respect to the {\it
  IUE}-based relations (IRX decreases because $L_{\rm FUV}$ increases)
and essentially eliminate the offset between starbursting and normal
star-forming galaxies (e.g., grey squares compared to cyan triangles
in the left panel of Fig.\ 3 of \citealt{overzier11}). The {\it IUE}
flux loss does not affect much the measurement of the UV slope or the
UV color. According to \citealt{takeuchi12}, the change in $\beta$
measured in a small aperture versus a large aperture, is $\sim
0.1$. In summary, measuring starbursts and normal star-forming
galaxies in a consistent way removes the offset, which we confirm here
(Fig.\ \ref{fig:age}, middle row). The use of the relations
based on {\it IUE} measurements, primarily the \citet{meurer99} and
\citet{kong04} relations with modern UV measurements should be avoided.

Is it nevertheless possible that just the central starbursts follow a
different relation from the one based on integrated UV measurements?
To answer this question one would have to measure the IR luminosity in the
same small aperture as the UV flux, which {\it IRAS} could not
do. This was made possible to some extent by \citet{takeuchi12},
using the {\it AKARI} IR data. The results (their Figs.\ 10 and 11)
demonstrate that there is not much difference in the IRX-$\beta$ locus
of the central and the integrated measurements of starbursts, thus
preserving the similarity of starbursts and normal galaxies in terms
of their IRX-$\beta$, and, consequently, their attenuation curves.

\subsection{Dust attenuation curve as the driver of the IRX-$\beta$
  scatter that itself depends on other factors}

The result that the scatter in the IRX-$\beta$ relation is driven
entirely by the diversity of attenuation curves may seem at odds with
numerous studies that have discussed, or have identified, other
factors, or a combination of factors that may or may not include the
attenuation curve. This difference is only apparent. Our analysis
shows that those ``other'' factors may have their affect on the
IRX-$\beta$ relation, but fundamentally {\it through} their effect on
the attenuation curve. Therefore, the IRX-$\beta$ scatter is simply
the manifestations of the differences in attenuation curves,
facilitated by the fact that the intrinsic UV slopes ($\beta_0$) of
star-forming galaxies span a small range, much smaller than the range
of observed UV slopes (upper left panel of Fig.\ \ref{fig:other} and
Fig.\ \ref{fig:ssfr_beta0}). The narrow range of intrinsic UV slopes
is expected in relatively continuous SF histories
\citep{leitherer99,calzetti01} that characterize integrated galaxy
light. Indeed, we find that the models predict that the range of
$\beta_0$ remains narrow even if there is a strong, quickly decaying
recent burst, as long as it lies on top of a non-negligible continuous
SFR, which, in the local universe and for integrated light of galaxies
is always present. We have redone the entire analysis with an expanded
model grid that includes young, quickly decaying bursts and the
results have remained unchanged: the scatter in the IRX-$\beta$ plane
vanishes once the attenuation curve is fixed, because the range of
$\beta_0$ stays narrow.

Our results may not hold to individual star-forming regions within
galaxies (e.g.,
\citealt{gordon04,calzetti05,thilker07,munozmateos09,boquien09,boquien12})
where the star-formation histories may be more similar to an
instantaneous burst that plummets to zero SFR. How similar the SF
history of a SF region may be to an instantaneous burst probably also
depends on how small the measured region is. In cases where the SF
ceases completely after a brief burst, the effects of aging will lead
to a significant reddening of both the intrinsic UV slopes
\citep{leitherer99,calzetti01} and also of the UV-optical
colors. Furthermore, the effects of dust-star geometry may be
different for SF regions than when considering entire galaxies,
complicating the interpretation.

Returning to integrated light og galaxies, we illustrate the ultimate
role of the attenuation curves by the following example. Several
studies have pointed out the principal role of dust grain distribution
(i.e., the dust type), in driving the IRX-$\beta$ scatter (e.g.,
\citealt{panuzzo07,popping17,safarzadeh17}). The attenuation curve is
the end result of the intrinsic {\it extinction} curve, which depends
on the dust properties \citep{weingartner01}, and of the relative
distribution of dust and stars of one or multiple populations (the
{\it geometry}). We do not know much about the extinction curves (dust
types) of galaxies other than the ones where we can probe the line of
sight to individual reddened stars (i.e., the MW, LMC and SMC), and
even in these galaxies there is much variation along different lines
of sight. Furthermore, even the fixed extinction curve will result in
very different {\it attenuation curves} depending on the dust-star
geometry (e.g., \citealt{panuzzo07,popping17,narayanan18}). However,
the point we wish to make is that how the resulting attenuation curve
arose (different dust type or geometry) does not change the fact that
there is a very close connection between the resulting attenuation
curve and the IRX-$\beta$ locus. Conceptually speaking, the role of
the geometry and the dust types are the questions that pertain
directly to the attenuation curve, and only by extension to
IRX-$\beta$. The question of the diversity of attenuation curves is
more fundamental that its manifestation in IRX-$\beta$, both because
the former is more general, as it pertains to a broader wavelength
range, and also because in the UV region the effect of changing the
attenuation curve slope and changing the UV bump strength will be
somewhat degenerate (see also \citealt{mao14}). Thus it is preferable
to investigate the questions relating to the diversity of attenuation
curves directly, rather than via IRX-$\beta$, which should primarily
be used as a tool to estimate FUV dust correction.

What can we say about the drivers of the variation in attenuation
curves that manifests as IRX-$\beta$ scatter? Of the parameters that
we studied here, it is the optical opacity ($A_V$) that matters the
the most. Other, weaker, trends (with age, sSFR or stellar mass)
vanish once $A_V$ is fixed, demonstrating that these trends not
primary but operate via their correlation with $A_V$. That the optical
opacity is strongly related to dust attenuation slope has been shown
using GSWLC-M2 dataset in \citet{s18}, and was also obtained as a
result in radiative transfer models
\citep{witt00,pierini04,chevallard13,seon16,narayanan18}. It appears
to be the consequence of the changing contributions of scattering
(which is strongly wavelength dependent) and absorption (which is
not), the former being dominant in low-opacity galaxies. The
relationship between the dust attenuation curve and $A_V$ does have an
intrinsic scatter, since, at the minimum there must be galaxy to
galaxy differences in dust type. It appears that for a given dust mass
the geometry, including the turbulence \citep{seon16} or the covering
fraction of dust, just change $A_V$ \citep{popping17}, so they again
may not represent independent factors that determine the shape of the
attenuation curve. Similarly, galaxy inclination may increase the
observed $A_V$ from its intrinsic (face-on) value, but the resulting
relationship between observed $A_V$ and the attenuation curve will be
the same as for the face-on galaxy with that higher value being its
intrinsic $A_V$ \citep{chevallard13,s18}. In other words, galaxies
with different ages, of varying amount of turbulence, inclination,
etc.\ affect attenuation curve and IRX-$\beta$ scatter through $A_V$
and not independently from it.


%
\subsection{IRX-$\beta$ relation at higher redshifts}

Most low-redshift studies that explore the role of the attenuation
curve in driving the IRX-$\beta$ scatter, consider it as one of the
posible factor whose significance with respect to other factors needs
to be established (e.g., \citealt{burgarella05,boquien12,mao14}). We
started by following such approach in this study, but after
establishing that the IRX-$\beta$ scatter is simply the manifestation
of the diversity of attenuation curves, we switched to considering
what other factors affect both the IRX-$\beta$ and the attenuation
curve. Interestingly, in many high-redshift studies the close
connection between the attenuation law and the IRX-$\beta$ is taken
for granted, and the IRX-$\beta$ is then naturally considered as a
tool to learn about the dust attenuation curve (\citealt{siana09} as
an early example of such an approach). For example, \citet{salmon16}
present relatively narrow loci of model points in the IRX-$\beta$
plane corresponding to different attenuation and extinction
curves. Different tracks clearly cover a wide range in the IRX-$\beta$
plane, in itself suggesting the dominant role of attenuation curves in
driving the scatter. Indeed, the authors consider these tracks to be
the ``observational basis'' for constraining the attenuation
curves. Similar attitude is present in other high-redshift studies
(e.g., \citealt{lofaro17,reddy18,mclure18}). \citet{cullen17}
explicitly state that for a constant $\beta_0$ there exists a simple
mapping between the effective attenuation curve and the
$A_{\rm FUV}$-$\beta$ relation.

\begin{figure}
\epsscale{1.2} \plotone{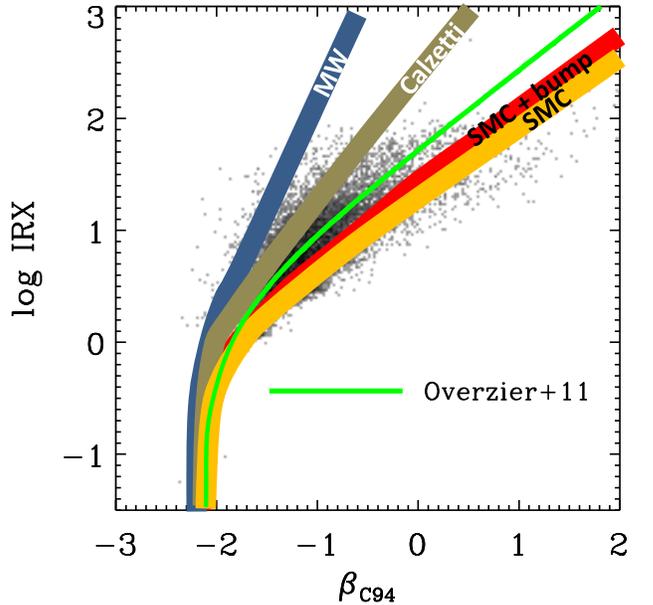}
\caption{IRX-$\beta$ for our sample superimposed with the loci
  expected for several attenuation/extinction curves. Predicted tracks
  are narrow, in line with our findings that attenuation curves drive
  the IRX-$\beta$ scatter. Milky Way and SMC extinction curves are
  treated as attenuation curves. Average galaxy agrees with a curve
  that is almost as steep as the SMC, but with some UV bump. At $z>1$,
  the tracks will be shifted to the left by up to several tenths, to
  reflect intrinsically steeper UV curves of very young
  galaxies. \label{fig:dal}}
\end{figure}

When high-redshift studies have concerns regarding the close
connection between the IRX-$\beta$ plot and the attenuation curve, it
is primarily because of the possible uncertainty regarding the values of
the intrinsic UV slopes of young galaxies \citep{buat12}, since the
form of the correspondence between the IRX-$\beta$ point distribution
and attenuation curves relies on knowing the intrinsic UV
slope. \citet{reddy18} argue that the typical intrinsic UV slope at
$z\sim2$ is $\beta_0=-2.62$, in line with the predictions of
BPASS models \citep{stanway16}, in contrast to what is assumed (and
measured) at low redshift, where $\beta_0$ does not go below $-2.3$ even for
extreme metal-poor and essentially dust-free starbursts such as I Zw
18 \citep{gildepaz07}, and is typically much higher (Fig.\
\ref{fig:other}, upper left). Our nominal models based on \citet{bc03}
allow $\beta_{0,{\rm min}}=-2.56$ for low metallicity objects with
recent bursts, but very few objects, including local starbursts,
approach such low value locally. Our preliminary analysis of the
CANDELS sample at $z\sim 2$ yields median $\beta_0=-2.36$, just 0.12
bluer than locally. Simulated galaxies at $z\sim 5$ have a similar
average $\beta_0=-2.4$, with the small scatter of 0.05 \citep{cullen17}.

In light of the view of a close connection between the IRX-$\beta$
point distribution and the attenuation curve, we present the
IRX-$\beta$ diagram on which we superimpose the loci corresponding to
several extinction/attenuation curves (Figure
\ref{fig:dal}). Specifically, we show the track corresponding to the
MW extinction curve (\citealt{cardelli89} with \citealt{odonnell94}
update), the \citet{calzetti00} curve, and the SMC \citep{prevot84}
curve. Even though the MW and the Calzetti curves have similar overall
slopes (in $A_{\lambda}/A_V$ formulation), the presence of a strong
bump in the MW curve results in a different locus in the IRX-$\beta$
with respect to the Calzetti curve. We have determined the location of
the tracks by forcing the above curves in our SED fitting with CIGALE
and appropriate modifications to the dust module. We implemented the
attenuation curves in two ways, treating each curve
as an effective attenuation curve, and as intrinsic attenuation curves
in in age-dependent implementation (the adopted method in this paper).
The two application results in slight shifts,
(especially for the MW curve) which are encompassed by the width of
the colored lines in Figure \ref{fig:dal}. Note that some papers
differ in the placement of these curves. In particular, the SMC track
is sometimes shown to lie below our SMC track. The observed
IRX-$\beta$ values in the local universe span the range of these
fiducial curves, in broad agreement with the analysis of attenuation
curves in \citet{s18}. It is worth noting that the inferences from
interpreting IRX-$\beta$ in terms of canonical {\it extinction} curves
are limited, since we do not expect attenuation curves to look like
Milky Way's or SMC's {\it extinction} curve, except by chance. Indeed,
the results from \citet{s18} suggests that attenuation curves that
look like MW (shallow strong bump) or SMC (steep, no bump) curves are
not typical.  In \citet{s18} we have shown that a typical curve of a
low-redshift galaxy has a slope almost as steep as the SMC slope, but
also featuring a modest bump. Indeed, if we were to add a bump to the
SMC curve (with strength $B=2.5$), SMC's locus would shift towards the
\citet{overzier11} relation and closer to the middle of the observed
points. This shift due to the bump, albeit small, emphasizes the point
that direct determination of an attenuation curve from multiband
photometry that straddles the bump will be more powerful than deriving
the attenuation curve from the IRX-$\beta$. The latter is subject to
the degeneracy introduced by the presence of the UV bump, which
affects even the $\beta_{\rm C94}$ slope designed to minimize the
influence of the bump region.

The tracks shown in Figure \ref{fig:dal}) are calculated for our local
sample. Any evolution towards bluer intrinsic UV slopes at higher
redshifts will be reflected in the shift of these curves to the left,
by $\Delta \beta_0$, which is currently uncertain, but is probably no
more than a few tenths. In such case, the form of the correspondence
between the IRX-$\beta$ and the attenuation curves will mostly be
similar.


\section{Summary} \label{sec:conc}

We summarize our findings as follows:

\begin{enumerate}

\item The general population of low-redshift galaxies exhibits a
  relatively large scatter in the IRX-$\beta$ plane.
\item The scatter in the IRX-$\beta$ diagram of star-forming galaxies is
  driven entirely by the diversity of dust attenuation curves
  (primarily their slopes, and to lesser degree the varying strength of
  their UV bumps).
\item Consequently, the question of the IRX-$\beta$ scatter of
  star-fromning galaxies becomes derivative of the question of the
  variations in the shapes of dust attenuation curves. The latter
  appears to primarily depend on the optical opacity ($A_V$, \citealt{seon16,s18}), and
  presumably the intrinsic variations in dust type, i.e., grain size
  distribution \citep{weingartner01}.
\item The fundamental reason why the dust attenuation curve is the
  driver of the scatter in the IRX-$\beta$ diagram is because star-forming
  galaxies (log sSFR$>-10.5$) have a relatively small range of
  intrinsic (dust-free) UV slopes (90\% are within 0.1 of $\beta=-2.24$).
\item The position of a star-forming galaxy on the IRX-$\beta$ diagram
  is the result only of the shape of the dust attenuation curve and
  dust opacity, all other factors being indirect, i.e., affecting
  these two.
\item Galaxies of different population ages or different sSFR
  (whether absolute, or relative to the main sequence), have, on
  average, very similar dust attenuation curves and, consequently,
  similar IRX-$\beta$ distributions.
\item Similarly, the starbursting galaxies and normal star-forming
  galaxies exhibit no significant shift in average IRX-$\beta$
  relations, and are both well described by \citet{overzier11}
  relation. Previously reported shift can be explained by the known,
  factor of 2--3, aperture loss of the FUV flux that affects the
  commonly-used local starburst relations derived from $\it IUE$ data.
\item Passive galaxies (log sSFR$<-11$) have much redder intrinsic UV
  slopes ($-2<\beta_0<0$) than star-forming galaxies, which are
  furthermore strongly dependent on the stellar
  metallicity. Nevertheless, the IRX-$\beta$ scatter for passive
  galaxies of fixed dust attenuation curve slope is not much increased
  with respect to the star-forming galaxies due to a fortuitous fact
  that for solar-like metallicities the SF history vector and the
  attenuation vector are parallel. Therefore, the dust attenuation
  curve effectively remains the principal driver of the IRX-$\beta$
  scatter for passive galaxies as well.
\item As a consequence of the above points, the IRX-$\beta$
  diagram can be used as a direct tool to determine the slope of the
  attenuation curve for star-forming galaxies in the UV region (e.g.,
  using Eq. \ref{eqn:deltaeta}). However, the UV slope $\beta$ suffers
  from some degeneracy between the attenuation curve slope and UV bump
  strength, so it is preferable and more comprehensive to base
  attenuation curve determinations on the full SED fitting, especially
  the one that includes IR SED or IR luminosity as a constraint.
\item The placement of attenuation curve tracks on the IRX-$\beta$
  diagram will shift to the left for galaxies at high redshift ($1<z<5$), to
  account for their bluer intrinsic UV slopes, by the value that is
  currently uncertain, but may be as low as $\Delta \beta=-0.1$
  (a negligible change) or as high as $-0.4$ (a moderate change).
\item Ignoring the scatter in the IRX-$\beta$ relation by using a
  single, fixed relation to correct UV SFRs introduces a random error
  of 0.15 dex to be added in quadrature to other sources of SFR
  errors. 
\item We provide relations that allow the IRX-$\beta$ relation to be
  moderately improved (20\% reduction in variance) by including the
  dependence on the stellar mass \citep{alvarez16}, or on the Balmer
  decrement. The IRX-$M_*$ relation alone is considerably inferior to
  the IRX-$\beta$ relation (0.7 mag of scatter in $A_{\rm FUV}$ vs.\
  0.4 mag).
\end{enumerate}

While many studies have explored the role of attenuation curve on the
IRX-$\beta$ relation, our principal contribution is in showing that
the two aspects, when considering entire galaxies, are fundamentally
connected owing to a very small range of intrinsic UV slopes in
star-forming galaxies. This result should help the field more
efficiently focus on the more fundamental question, what shapes the
dust attenuation curve, for which significant advances, both
observational and theoretical, are being made. To that end, the new
IRX-$\beta$ parameterization that we introduce in this paper, which
can be used to directly estimate the slope of the attenuation curve,
can also be useful.

\acknowledgments We are grateful to V\'eronique Buat and the reviewer
for many helpful and thought-provoking comments. The construction of
GSWLC was funded through NASA ADAP award NNX12AE06G. MB acknowledges
support from FONDECYT regular grant 1170618. Funding for SDSS-III has
been provided by the Alfred P. Sloan Foundation, the Participating
Institutions, the National Science Foundation, and the U.S. Department
of Energy Office of Science. The SDSS-III web site is
http://www.sdss3.org/. Based on observations made with the NASA Galaxy
Evolution Explorer. GALEX is operated for NASA by the California
Institute of Technology under NASA contract NAS5-98034. This
publication makes use of data products from the Wide-field Infrared
Survey Explorer, which is a joint project of the University of
California, Los Angeles, and the Jet Propulsion Laboratory/California
Institute of Technology, funded by the National Aeronautics and Space
Administration.


\end{document}